\documentclass[11pt,letterpaper]{article}
    \usepackage{jheppub}
\usepackage{amsmath}
\usepackage{amssymb}
\usepackage{bm}
\usepackage{graphicx}
\usepackage{dsfont}

\newcommand{\Tr}{\mathrm{Tr} \,}
\newcommand{\SU}{\mathrm{SU}}
\newcommand{\SUn}{\mathrm{SU}(n)}

\newcommand{\T}{\mathcal{T}}
\newcommand{\R}{\mathcal{R}}
\newcommand{\mlight}{m_{\rm light}}
\newcommand{\mbar}{\overline{m}}
\newcommand{\Fhat}{\widehat{F}}
\newcommand{\Ahat}{\widehat{A}}
\newcommand{\hop}{\text{hop}}
\newcommand{\NL}{\text{NL}}
\newcommand{\be}{\begin{equation}}
\newcommand{\ee}{\end{equation}}
\newcommand{\Ref}[1]{Ref.~\cite{#1}}
\newcommand{\Refs}[1]{Ref.~\cite{#1}}
\newcommand{\Fig}[1]{Figure~\ref{#1}}
\newcommand{\Figs}[2]{Figures~\ref{#1} and \ref{#2}}
\newcommand{\Sec}[1]{Section~\ref{#1}}
\newcommand{\App}[1]{Appendix~\ref{#1}}
\newcommand{\Eq}[1]{Eq.~(\ref{#1})}
\newcommand{\Eqs}[2]{Eqs.~(\ref{#1}) and (\ref{#2})}
\newcommand{\eq}[1]{(\ref{#1})}

\newcommand{\one}{\mathds{1}}

\title{Locality in Theory Space}
\author{Yonatan Kahn}
\author{and Jesse Thaler}
\affiliation{Center for Theoretical Physics,\\ Massachusetts Institute of Technology, Cambridge, MA 02139, U.S.A.}
\emailAdd{ykahn@mit.edu}
\emailAdd{jthaler@mit.edu}
                                           
\abstract{Locality is a guiding principle for constructing realistic quantum field theories.  Compactified theories offer an interesting context in which to think about locality, since interactions can be nonlocal in the compact directions while still being local in the extended ones.   In this paper, we study locality in ``theory space'', four-dimensional Lagrangians which are dimensional deconstructions of  five-dimensional Yang-Mills.  
In explicit ultraviolet (UV) completions, one can understand the origin of theory space locality by the irrelevance of nonlocal operators.  From an infrared (IR) point of view, though, theory space locality does not appear to be a special property, since the lowest-lying Kaluza-Klein (KK) modes are simply described by a gauged nonlinear sigma model, and locality imposes seemingly arbitrary constraints on the KK spectrum and interactions.  We argue that these constraints are nevertheless important from an IR perspective, since they affect the four-dimensional cutoff of the theory where high energy scattering hits strong coupling.   Intriguingly, we find that maximizing this cutoff scale implies five-dimensional locality.  In this way, theory space locality is correlated with weak coupling in the IR, independent of UV considerations.  We briefly comment on other scenarios where maximizing the cutoff scale yields interesting physics, including theory space descriptions of QCD and deconstructions of anti-de Sitter space.
}
                                           
\begin{document}

\hfill MIT-CTP 4346

\maketitle

\section{Introduction}
\label{sec:Intro}

Locality is a fundamental guiding principle when constructing quantum field theories to describe physical systems.  Locality appears in many different guises, from the causal structure of Lorentz-invariant theories to the analyticity of the $S$-matrix.  Theories with compact dimensions offer an interesting context in which to think about locality, since for a low-energy observer, locality in the compact dimensions is qualitatively different from locality in the noncompact ones. From an ultraviolet (UV) or top-down perspective, various mechanisms exist to ensure compact locality.  In the usual picture of dimensional reduction, locality in the UV is assumed, and interactions in the compact dimensions remain local after geometric compactification.  In models of dimensional deconstruction \cite{ArkaniHamed:2001ca}, a UV-complete four-dimensional gauge theory condenses at low energies to yield a theory with a compact fifth dimension, and five-dimensional locality is ensured by the irrelevance of nonlocal operators before condensation.  A deeper mechanism exists in the AdS/CFT correspondence \cite{Maldacena:1997re, Gubser:1998bc, Witten:1998qj}, where bulk locality emerges from the large-$n$ limit of the boundary CFT \cite{Heemskerk:2009pn, Heemskerk:2010ty, Fitzpatrick:2010zm, Sundrum:2011ic, Fitzpatrick:2011hu}.

From an infrared (IR) or bottom-up perspective, however, compact locality is baffling.  In the far IR, a compact dimension can be described by a tower of Kaluza-Klein (KK) modes, and locality simply enforces certain constraints on the spectrum and interactions of these modes.  If there are spin-1 degrees of freedom, as will be the case in this paper, there is a cutoff scale $\Lambda$ where longitudinal scattering of the massive spin-1 KK modes becomes strongly coupled.  From an IR point of view, there is no apparent reason to exclude additional nonlocal interactions, and one might even expect nonlocal terms could render the theory better behaved in the IR.   Indeed, in the local case, it is precisely the interactions among different KK levels which partially unitarize KK scattering, pushing $\Lambda$ above the naive expectation from considering the KK modes as independent massive vectors.  It is therefore plausible that including nonlocal interactions with the correct sign could yield a similar interference effect, possibly driving the cutoff scale $\Lambda$ higher than in the local case.

\begin{figure}[t]
\begin{center}
\includegraphics[scale=0.45]{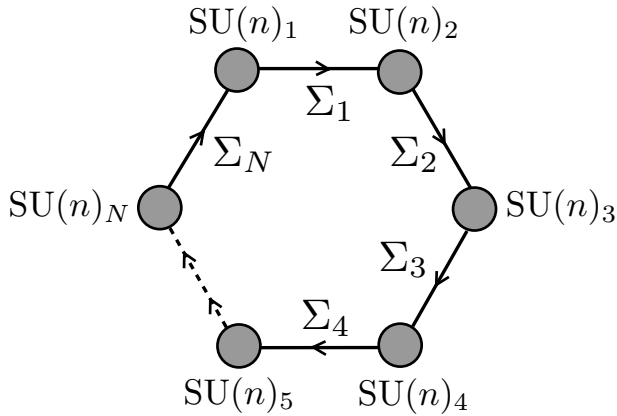}
\caption{Local cyclic $N$-site $\SUn$ moose diagram, known also as ``theory space,'' corresponding to the latticization of compactified five-dimensional Yang-Mills. The link fields $\Sigma$ transform as bifundamentals of the gauge groups, represented by shaded circles at either end of the link.  When each link field acquires a vacuum expectation value, the moose describes $N-1$ interacting massive $\SUn$ gauge bosons, one massless $\SUn$ gauge boson, and one Goldstone winding mode.  Examples of nonlocal interactions are shown in \Fig{fig:NLMoose}.}
\label{fig:CyclicMoose}
\end{center}
\end{figure}

In this paper, we present a system where precisely the opposite is true:  insisting on the highest possible cutoff scale $\Lambda$ \emph{implies} locality in the compact dimension.  We study the case of a deconstructed five-dimensional $\SU(2)$ Yang-Mills theory in a flat geometry, described by a ``theory space'' cyclic moose diagram as in \Fig{fig:CyclicMoose}.  This four-dimensional theory has an intrinsic cutoff scale $\Lambda$, and maximizing $\Lambda$ is correlated with locality in theory space.  This gives a purely low-energy perspective on why compact locality is special, in the sense that local theories are the most weakly coupled in the IR. Strictly speaking, our analysis only holds for small nonlocal perturbations, and we cannot exclude the possibility that large nonlocal terms could lead to a larger value of $\Lambda$.   While unitarity violation in higher-dimensional gauge theories has been investigated before \cite{SekharChivukula:2001hz,Chivukula:2002ej,DeCurtis:2003zt}, to our knowledge the only studies of extra-dimensional nonlocality have been in a gravitational context \cite{Schwartz:2003vj}.\footnote{When discretizing gravity, nonlocal interactions are necessary to have a local continuum limit \cite{Schwartz:2003vj}, which is not the case for gauge theories.} 

More concretely, in order to isolate the effects of nonlocality and remove trivial rescalings of $\Lambda$, we introduce a dimensionless ratio
\be
\label{eq:RATIO}
\R = \frac{\Lambda}{\mbar}
\ee
which normalizes the cutoff scale, where $\mbar$ is an average mass for the spin-1 modes whose precise definition will be given in \Sec{sec:Diagnostic}.  After adding a nonlocal gauge kinetic term with small coefficient $\epsilon$ and nonlocal length scale $\ell$, we show that $\R$ behaves to lowest order in $\epsilon$ as
\be
\label{eq:RATIOwithNL}
\R_{\NL} = \R_{\rm{local}}\left(1 - 4 \epsilon^2 N^2 \frac{\ell^2}{R^2} \right ), \qquad \ell < R,
\ee
where $R$ is the compactification radius and $N$ is the number of sites in the moose. Surprisingly, terms which would contribute to scattering amplitudes at linear order in $\epsilon$ are (miraculously) absent due to a group-theoretic cancellation, thanks to the fact that the dominant scattering channel is a gauge singlet. This results in a leading $\epsilon$ dependence in \Eq{eq:RATIOwithNL} which is quadratic, such that locality at $\epsilon = 0$ is a special value. We specialize to $\SU(2)$ for technical reasons, but the result that maximizing $\R$ implies locality holds any gauge group where the singlet channel dominates the scattering matrix; in particular, it holds for general $\SUn$, up to possible corrections that are subleading in $N$.

Beyond cyclic theory space, the ratio $\R$ is interesting in at least two other settings.  First, as is well known, the pions and $\rho$ mesons of quantum chromodynamics (QCD) can be described by a three-site moose, but \Ref{Georgi:1989xy} showed that ``nonlocal'' interactions with a negative nonlocal coefficient are required to match known QCD phenomenology.  We find that maximing $\R$ does indeed favor a nonlocal interaction, but one whose sign depends on the choice of normalization factor $\mbar$.  Second, when deconstructing spaces with nontrivial geometry, such as warped anti-de Sitter (AdS) geometries in Randall-Sundrum scenarios \cite{Randall:1999ee, Randall:1999vf}, maximizing $\R$ implies an ``$f$-flat deconstruction'' with all link decay constants equal. Such a deconstruction was first noted in \Ref{Abe:2002rj}, and used in \Ref{SekharChivukula:2006we}  as a convenient simplification for sum rule computations, but here we show that it leads to a deconstructed theory with a maximal domain of validity (preferable to the usual deconstruction of AdS with equal lattice spacings \cite{Randall:2002qr,Falkowski:2002cm}). Furthermore, the same gauge singlet channel dominates even in warped geometries, giving circumstantial evidence that maximizing the cutoff also implies locality in AdS$_5$. 

The remainder of this paper is organized as follows. In \Sec{sec:Notation}, we set the notation and conventions for the moose diagram which describes the deconstructed theory, and define the nonlocal perturbations in \Eq{CyclicONL}.  In \Sec{sec:Diagnostic}, we motivate the form of $\R$ in \Eq{eq:RATIO} using dimensional and scaling arguments to define the average mass scale $\mbar$ in \Eq{DeconstructedMbar}.   In \Sec{sec:TMatrixLocal}, we describe the coupled-channel analysis of gauge boson scattering to define the cutoff scale $\Lambda$ and carry out the calculation of $\R$ for the local moose.  We derive our main result in \Sec{sec:NSiteNL}, showing that maximizing $\R$ implies locality in a cyclic $\SU(2)$ $N$-site moose with nonlocal terms.  In \Sec{sec:OtherUses}, we briefly illustrate the phenomenological consequences of our ratio $\R$ for a three-site QCD moose and for warped deconstructions, and speculate about locality in AdS space.  We summarize our results in \Sec{sec:Conclusion} and suggest ways to extend our analysis beyond tree-level.

\section{Moose notation for theory space}
\label{sec:Notation}

In the following two subsections, we define our notation for the moose diagram representing theory space, and review the phenomenology of dimensional deconstruction. A reader familiar with these results may wish to skip to \Sec{sec:NLTerms} where we introduce nonlocal interactions in theory space.

\subsection{Local Lagrangian}

We define a local $N$-site \emph{cyclic moose} with gauge group $\SUn$ by the Lagrangian
\begin{equation}
\label{CyclicMooseLag}
\mathcal{L}_{\rm cyc} = -\frac{1}{2} \sum_{i=1}^N \Tr F_i^2 +  \sum_{j=1}^{N}f_j^2 \, \Tr |D_\mu \Sigma_j|^2,
\end{equation}
as shown in \Fig{fig:CyclicMoose}.  The $\mathfrak{su}(n)$-valued field strengths $F^i_{\mu \nu}(x)$ live on the sites $i = 1, 2, \dots, N$, and the link fields $\Sigma_j \in \SUn$ live on the links between sites $j$ and $j+1$.\footnote{For ease of notation, we freely raise and lower site indices on gauge and link fields.} There is a periodic identification of sites and links by $i \simeq i + N$. The sites are shaded to represent gauge symmetries, as opposed to global symmetries which will be important in our discussion of the QCD moose in \Sec{sec:Threesite}. 

Our normalization for the $\mathfrak{su}(n)$ generators $T^a$ is
\begin{equation}
\label{SUnGenNorm}
\Tr(T^a T^b) = \frac{1}{2} \delta^{ab}.
\end{equation}
The gauge transformations of the link fields are
\begin{equation}
\Sigma_j \to U_j^\dagger(x) \Sigma_j U_{j+1}(x),
\end{equation}
where $U_{j}, U_{j+1} \in \SUn$ are in the fundamental representation, so $\Sigma_j$ transforms in the antifundamental $\overline{\mathbf{n}}$ of the site to its left and the fundamental $\mathbf{n}$ of the site to its right. This behavior is represented by the directional arrows on the link fields in \Fig{fig:CyclicMoose}. The covariant derivative is defined by
\begin{equation}
\label{CovDerivSigma}
D_\mu \Sigma_j = \partial_\mu \Sigma_j - ig_j A^j_\mu \Sigma_j + ig_{j+1} \Sigma_j A^{j+1}_\mu,
\end{equation}
where we use canonical normalization for the gauge fields so the gauge couplings $g_j$ appear explicitly.

Writing
\begin{equation}
\label{SigmaPi}
\Sigma_j(x) = e^{i \pi_j^a(x) T^a /f_j},
\end{equation}
the Lagrangian \eq{CyclicMooseLag} becomes a nonlinear sigma model in 3+1 dimensions describing the interactions of the ``pions'' $\pi_j$ with themselves and the gauge fields $A^i_\mu$. Pursuing this analogy with low-energy QCD, we refer to $f_j$ as \emph{decay constants}. With the normalization convention \eq{SUnGenNorm}, we obtain canonically normalized kinetic terms for the $\pi$ fields after expanding the Lagrangian as a power series in the $\pi_j^a$. The remainder of the Lagrangian involving $\pi_j^a$ consists of derivative interactions suppressed by powers of $1/f_j$, all of which are nonrenormalizable operators.  Thus, the nonlinear sigma model has some UV cutoff, which we will take to be the scale of tree-level unitarity violation $\Lambda$; we will define $\Lambda$ precisely in \Sec{sec:PartialWave}. Keeping only the leading terms with coefficients $1/f_j^2$, one can estimate the scale of unitarity violation by ``naive dimensional analysis'' \cite{Manohar:1983md}, 
\be
\label{NDAbound}
\Lambda \sim 4 \pi \, \min_j \{ f_j \}.
\ee
Note that $\Lambda$ is determined by the minimum of the $f_j$ because $\pi_i$ and $\pi_j$ are decoupled for $i \neq j$, so each link has its own scale of unitarity violation, and unitarity violation for the whole moose is dominated by whichever one occurs first.

With the parameterization \eq{SigmaPi}, each $\Sigma_j$ gets a vacuum expectation value (vev) \linebreak $\langle \Sigma_j \rangle=  \one$ and the link fields spontaneously break the gauge symmetry at each site, with $\pi_j$ acting as Goldstone bosons which are eaten by the gauge fields $A^j_\mu$. The remaining unbroken gauge symmetry is the diagonal $\SUn$ subgroup, whose gauge coupling $g_4$ (in a notation suggestive of KK decomposition) is given by 
\be
\label{g4Moose}
\frac{1}{g_4^2} = \sum_{i=1}^N \frac{1}{g_i^2}.
\ee
Since this is a cyclic moose, there is also an uneaten linear combination of Goldstone modes $\tilde{\pi}= (\pi_1 + \pi_2 + \cdots + \pi_N)/\sqrt{N}$, which we will sometimes refer to as the \emph{winding mode.}

The mass-squared matrix for the now-massive gauge fields arises from the second term in \Eq{CyclicMooseLag} by setting $\Sigma_j = \langle \Sigma_j \rangle = \one$.  Setting the gauge couplings to a common value $g$,
\begin{equation}
\label{CyclicMooseMass}
M^2  =  g^2 \begin{pmatrix}
f_N^2+f_1^2 & -f_1^2 & \ 0 & \cdots & -f_N^2 \\
-f_1^2 & f_1^2 + f_2^2 & \ -f_2^2 & \cdots & 0 \\
0 & -f_2^2 & \ \ddots &   &  \vdots \\
 \vdots &  &\ & f_{N-2}^2 + f_{N-1}^2 & -f_{N-1}^2   \\
  -f_N^2 &  & \ & -f_{N-1}^2 & f_{N-1}^2 + f_N^2   \\
\end{pmatrix}.
\end{equation}
This matrix has a zero eigenvalue for all choices of $f_j$, with eigenvector \linebreak $A^{(0)} = (A_1 + A_2 + \cdots + A_N)/\sqrt{N}$, corresponding to the diagonal subgroup mentioned above. If we further restrict the decay constants $f_j$ to be equal ($f_j \equiv f$), we have an analytic expression for the mass spectrum of the cyclic moose:
\begin{equation}
\label{CyclicMassSpectrum}
M_k^2 = 4g^2 f^2 \sin^2 \left ( \frac{\pi k}{N} \right ), \qquad -N/2 < k \leq N/2.
\end{equation}
With the exception of \Sec{sec:Warped} where we explore warped spaces, we will always set the decay constants and gauge couplings equal, $f_j \equiv f$ and $g_i \equiv g$, corresponding to a discrete translation invariance along the moose.

\subsection{Five-dimensional interpretation}
\label{sec:Dictionary}

The Lagrangian \eq{CyclicMooseLag} can be interpreted as a (4+1)-dimensional lattice gauge theory where only the compact fifth dimension has been latticized, with the $\Sigma_j$ providing the fifth component of the five-dimensional gauge field in the continuum limit where the lattice spacing goes to zero. The continuum limit is ordinary five-dimensional Yang-Mills
\be
\label{eq:5dYM}
S = -\frac{1}{2}\int d^5 x \, \Tr (\Fhat_{MN}\Fhat^{MN}),
\ee
where $M,N = 0, 1, 2, 3, 5$ and
\be
\Fhat_{MN}^a = \partial_M \Ahat^a_N - \partial_N \Ahat^a_M + g_5 f^{abc} \Ahat^b_M \Ahat^c_M.
\ee
Using this interpretation, known as dimensional deconstruction, we obtain a dictionary between parameters in the Lagrangian and parameters in the latticized theory \cite{ArkaniHamed:2001ca}:
\begin{align}
\label{ParamDictionary}
\text{Lattice spacing}: a & = \frac{1}{gf} \\
\text{Circumference of fifth dimension}: R & = Na \\
\label{dictg5}
\text{Five-dimensional gauge coupling}: g_5 & = \sqrt{\frac{g}{f}} \\
\text{Effective four-dimensional gauge coupling}: g_4 & = \frac{g}{\sqrt{N}}
\end{align}
The last of these relations identifies the gauge coupling of the diagonal subgroup ($g_4$ in \Eq{g4Moose}) with the the effective four-dimensional gauge coupling of the KK zero mode of the five-dimensional theory after compactification.

The five-dimensional interpretation of the deconstruction can be confirmed in several ways. First, the dictionary preserves the usual relation between four- and five-dimensional gauge couplings after compactification:
\be
\label{4d5d}
\frac{1}{g_4^2} = \frac{R}{g_5^2}.
\ee
Second, we can examine the mass spectrum of \Eq{CyclicMassSpectrum} in the continuum limit, $N \to \infty$ and $a \to 0$ with $R$ fixed:
\begin{equation}
\label{MassSpecContinuum}
M_k  \approx 2\pi |k| /R,
\end{equation}
for small integers $|k|$. This is precisely the KK spectrum of modes on a circle of circumference $R$, as one would expect from compactification of a fifth dimension. In this picture, the uneaten linear combination of Goldstone bosons corresponds to the nontrivial Wilson loop around the compact extra dimension, hence the name ``winding mode'' for $\tilde{\pi}$.

In the framework of dimensional deconstruction, locality in the latticized dimension is built into \Eq{CyclicMooseLag} through locality in theory space. Indeed, the field strengths $F^i_{\mu \nu}$ are decoupled at different sites $i \neq j$, and the $\Sigma_j$ only couple to nearest-neighbor gauge fields $A^j_\mu$ and $A^{j+1}_\mu$ through \Eq{CovDerivSigma}.   In the continuum limit, $D_\mu \Sigma_j$ becomes a covariant derivative along the latticed direction, which is also local. In the original application of dimensional deconstruction, \Ref{ArkaniHamed:2001ca} derived these local interactions by starting with a moose with additional fermions charged under a ``color'' gauge group, which confines to give the $\Sigma$ fields as fermion bilinears. However, in our analysis we take \Eq{CyclicMooseLag} as our starting point, with the $\Sigma_j$ as ``fundamental'' fields rather than composite operators.

\subsection{Nonlocal terms}
\label{sec:NLTerms}

The aim of this paper is to study nonlocality in theory space, which we will incorporate by perturbing the local moose \eq{CyclicMooseLag} by a gauge-invariant operator $\epsilon f^2 \mathcal{O}$, where $\epsilon$ is small and dimensionless and $f^2$ is inserted for normalization.\footnote{While the squares of the decay constants $f_j^2$ must be positive for the $\pi_j$ kinetic terms to have the correct sign, there is no such restriction on the sign of $\epsilon$, although $\epsilon$ must be real.}  For $\mathcal{O}$ to be nonlocal, it must connect distant sites $i$ and $j$, and hence transform in the $\overline{\mathbf{n}}$ of $\SUn_i$ and the $\mathbf{n}$ of $\SUn_j$. We could simply define a new link field $\widetilde{\Sigma}$ connecting these sites, but we are interested in comparing theories with the same number of four-dimensional degrees of freedom. Moreover, the theory with this extra field has a pathological local limit, since as its decay constant $\tilde{f}$ goes to zero, $\widetilde{\Sigma}$ disappears but unitarity is violated immediately due to \Eq{NDAbound}.\footnote{Even if fields like $\widetilde{\Sigma}$ were present in the original Lagrangian, we could always decouple them with the plaquette operator $\mu^2 f^2 \, \Tr|\Sigma_i \Sigma_{i+1} \cdots \Sigma_j \widetilde{\Sigma}^\dagger|^2$.  As we take $\mu \to \infty$, $\widetilde{\Sigma}$ becomes massive and decouples from the low-energy spectrum.}

\begin{figure}[t]
\begin{center}
\includegraphics[scale=0.45]{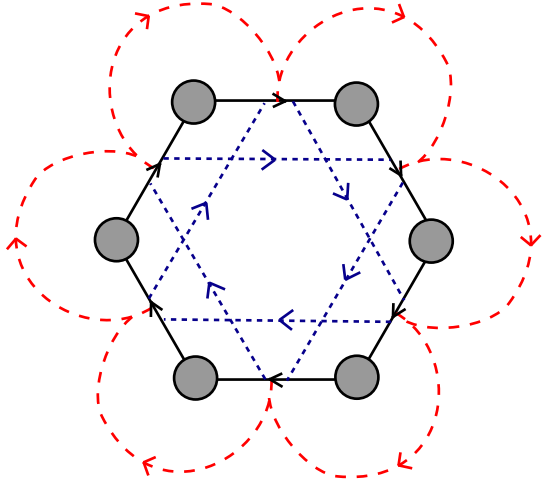}
\caption{A graphical representation of the nonlocal terms \eq{CyclicONL}, showing an $N = 6$ cyclic moose with nonlocal terms corresponding $N_{\hop} = 1$ (red long-dashed lines) and $N_{\hop} = 2$ (blue short-dashed lines).  We draw the nonlocal terms attached to links, rather than sites, to emphasize their construction in terms of link fields as in \Eq{Hopij}.}
\label{fig:NLMoose}
\end{center}
\end{figure}

Instead, we choose to consider
\begin{equation}
\label{Hopij}
\mathcal{O}_{N_{\hop}}^{(i)} \equiv 2 \, \Tr \left[ (D_\mu \Sigma_i) \Sigma_{i+1} \cdots \Sigma_{j} (D^\mu \Sigma_{j}^\dagger) \Sigma_{j-1}^\dagger \cdots \Sigma_i^\dagger \right]
\end{equation}
for $i < j$, and we will refer to this as a ``hopping'' term with $N_{\hop} = j-i$.\footnote{The factor of 2 is purely conventional and simplifies some formulas in what follows.} To preserve the discrete translation invariance in the compact dimension, we sum over all sites:
\begin{equation}
\label{CyclicONL}
\mathcal{O}_{N_{\hop}} \equiv \epsilon f^2 \sum_{i=1}^N \mathcal{O}_{N_{\hop}}^{(i)}.
\end{equation}
Figure \ref{fig:NLMoose} shows an example of a 6-site moose diagram with such nonlocal terms for both $N_{\hop} = 1$ and $N_{\hop} = 2$. The goal of \Sec{sec:NSiteNL} will be to study the effect of $\mathcal{O}_{N_{\hop}}$ on the cutoff scale $\Lambda$. For simplicity, we will only consider perturbing the local moose by a nonlocal term with a \emph{single} value of $N_{\hop}$ (unlike in \Fig{fig:NLMoose}) so that we can study the effect of nonlocality as a function of the nonlocal length scale.  We will abbreviate $\mathcal{O}_{N_{\hop}} \equiv \mathcal{O}_{\NL}$ and display all results as a function of $N_{\hop}$.

To construct the continuum version of this operator, we define the characteristic nonlocal length scale $\ell$ by
\be
\ell \equiv aN_{\hop}.
\ee
The deconstructed fields $\Sigma$ and $A_\mu$ are related to the continuum five-dimensional gauge field $\widehat{A}_M$ by
\begin{align}
\label{FieldDictionary}
\Sigma_j(x) &= P\exp\left[ -ig_5 \int_{(j-1)a}^{ja} dy \, \widehat{A}_5 (x, y)\right], \\
A^j_\mu(x) &= \sqrt{a} \widehat{A}_\mu(x,aj), \qquad \mu = 0, 1, 2, 3,
\end{align}
where $y$ is the coordinate along the fifth dimension and $P$ is the path-ordering symbol.  These relations imply that in the continuum limit,
\be
\label{Nonlocal5d}
\mathcal{O}_{\NL} \to 2\epsilon \int dy \, \Tr \left[ \widehat{F}_{\mu 5}(x,y)W(y,y+\ell) \widehat{F}^{\mu 5}(x,y+\ell) W^\dagger(y,y+\ell) \right],
\ee
so our nonlocal operator is a nonlocal gauge kinetic term, connecting $\widehat{F}_{\mu 5}(x,y)$ and \linebreak $\widehat{F}^{\mu 5}(x,y+\ell)$ by means of a Wilson line along the fifth dimension,
\begin{equation}
W(y,y+\ell) = P\exp\left[-ig_5 \int_{y}^{y+\ell} dy' \, \widehat{A}_5 (x, y')\right].
\end{equation}

In the five-dimensional picture, \Eq{Hopij} is the natural nonlocal operator to consider, as compared to operators like $|D_\mu (\Sigma_i \, \Sigma_{i+1} \, \cdots \Sigma_{j})  |^2$.   The continuum analogue of the latter is $|D_\mu W(y,y+\ell)|^2$, which does not have any simple interpretation. Moreover, such an operator is redundant, because 
\be
\label{eq:choiceOfNonLocal}
|D_\mu (\Sigma_i \, \Sigma_{i+1} \, \cdots \Sigma_{j})  |^2 = |D_\mu \Sigma_i|^2 + |D_\mu \Sigma_{i+1}|^2 + \cdots + |D_\mu \Sigma_{j}|^2 +  \cdots,
\ee
so this operator already contains kinetic terms for $\Sigma_i$ through $\Sigma_j$.\footnote{This redundant nonlocal operator will appear in the three-site QCD moose in \Sec{sec:Threesite}, and in the discussion of KK matching in \App{app:KKtruncDeconstruct}.} As a result, to lowest order in $\epsilon$ the effect of $\epsilon |D_\mu (\Sigma_i \, \Sigma_{i+1} \, \cdots \Sigma_{j})  |^2$ would be to simply shift the decay constant $f^2 \to f^2(1 + \epsilon)$, a trivial change which can be absorbed by a redefinition of the $\pi_i$. In addition, after subtracting off the kinetic terms from $\epsilon |D_\mu (\Sigma_i \, \Sigma_{i+1} \, \cdots \Sigma_{j})  |^2$, we are left precisely with terms like $\mathcal{O}_{N_{\hop}}^{(i)}$, but for \emph{all} values of $N_\hop \leq |j-i|$.  To isolate a single nonlocal length scale $\ell$, we will only consider the nonlocal operator (\ref{CyclicONL}).

We are always free to add additional terms (local or nonlocal) which have more than two derivatives, as long as they respect the symmetries present in the original local Lagrangian.  However, in a momentum power-counting scheme, it is consistent to truncate the effective Lagrangian to order $p^2$ provided we restrict the analysis to tree-level, as we will do in this paper. Indeed, two derivatives are the minimum required to have any nontrivial nonlocal interaction, and these nonlocal terms in \Eq{CyclicONL} are the same order in momentum as the nonlinear sigma model kinetic terms $|D_\mu \Sigma|^2$ appearing in the local Lagrangian. If we wanted to study the effects of higher-derivative terms, then by Weinberg's power-counting theorem \cite{Weinberg:1978kz}, we would also have to consider one-loop amplitudes of two-derivative terms. Such effects are certainly important, but we leave an analysis to order $p^4$ to future work.\footnote{This power-counting also explains why we do not consider nonlocal terms like $\Tr (F_{\mu \nu} \, \Sigma \, F^{\mu \nu} \, \Sigma^\dagger)$; as is well-known from chiral perturbation theory, these terms count as $p^4$ in momentum.}

\section{Normalizing the cutoff scale}
\label{sec:Diagnostic}

We argued in the introduction that the cutoff scale $\Lambda$ gives useful information about the weak-coupling domain of validity of a theory.  Since $\Lambda$ can be raised or lowered by trivial rescaling of the mass scales of the problem, one would prefer to study a dimensionless ratio between the cutoff scale and some other physical mass scale $\mbar$.  In this section, we define the normalization $\mbar$ which appears in our dimensionless ratio,
\be
\R = \frac{\Lambda}{\mbar}.
\tag{\ref{eq:RATIO}}
\ee
The precise definition of $\Lambda$ will be given in \Sec{sec:TMatrixLocal}, but for the arguments in this section, the dimensional analysis estimate (\ref{NDAbound}) will suffice.

Our goal is to find a definition for $\mbar$ that depends only on physical observables and has a well-defined continuum limit.  We will use the freedom to define $\mbar$ such that $\R$ is sensitive mainly to the degree of nonlocality, and is as insensitive as possible to local aspects of the four-dimensional moose theory and its five-dimensional interpretation.  We will further insist that the effects of nonlocality show up only in $\Lambda$, not in $\mbar$, such that $\R$ truly tests how locality affects the UV cutoff alone, not on how locality affects the spectrum of the theory.   

One possibility is to simply take $\mbar = f$, the decay constant of the moose, but this is problematic for several reasons.  First, $f$ does not have a nice continuum interpretation, nor is $f$ well-defined in mooses with unequal decay constants.    More significantly, $f$ cannot be unambiguously defined in terms of a physical observable.  In the local theory, $f$ can be made physical by identifying it with the 4-pion scattering amplitude, but with the addition of nonlocal terms, there is no unambiguous definition of $f$, since new scattering channels open up which were not present in the original Lagrangian.\footnote{There is an unambiguous definition of the decay constant $\tilde{f}$ for the uneaten Goldstone mode $\tilde{\pi}$, but $\tilde{f}$ can be adjusted at will by adding a kinetic term for this mode, $c_0 \sum_{i=1}^N |D_\mu(\Sigma_i \Sigma_{i+1} \cdots \Sigma_1 \cdots \Sigma_{i-1})|^2$, which does not affect the gauge boson mass matrix.  Thus, $\tilde{f}$ does not represent a physically relevant mass scale in the problem, and is not suitable for a normalization.}

Another option is to use a mass scale derived from the mass matrix $M^2$ in \Eq{CyclicMooseMass}. We have to be a bit careful, since any such mass $\mbar$ is proportional to $g$.  If we take $g \to 0$, $\Lambda/\mbar$ would be sent to infinity, but this tells us nothing about locality as we have artificially squashed the whole mass spectrum far below the cutoff.  To correct for this, we should normalize by $g_4$.\footnote{We choose $g_4$ rather than $g$ in anticipation of \Sec{sec:Warped}, since $g_4$ is well-defined even for unequal $g_i$ through \Eq{g4Moose}.} A seemingly natural choice for the mass scale is $\mbar = \mlight/g_4$, the (normalized) lightest nonzero mass eigenvalue.  This is appealing since $\Lambda/\mlight$ is essentially the spacing between the highest and lowest mass scales in the theory. However, $\mlight$ is not terribly useful in isolating the effects of nonlocality on $\Lambda$, since the gauge boson mode corresponding to $\mlight$ is already ``nonlocal,''  with nonzero components at most sites even in the local moose. 

Instead, the choice for $\mbar$ that we will use is
\be
\label{DeconstructedMbar}
\mbar = m_{\rm avg} \equiv \frac{\sqrt{\Tr M^2}}{Ng_4},
\ee
which is a sort of ``average'' mass.  As we will show in \Sec{sec:NSiteNLMass}, this choice is appealing since it is independent of the nonlocal perturbation (up to discretization effects).  Here, we will show that $m_{\rm avg}$ also has a nice continuum limit.   Consider the following scaling of the (local) deconstructed parameters:
\be
\label{DeconstructedScaling}
f \to \alpha f, \hspace{4mm} g \to \alpha g, \hspace{4mm} N \to \alpha^2 N.
\ee

This scaling preserves all the continuum parameters $g_5$, $R$, and $g_4$ according to the deconstruction dictionary in \Sec{sec:Dictionary}, but changes $\Lambda \rightarrow \alpha \Lambda$ according to \Eq{NDAbound}. For arbitrary $\alpha > 1$, we can approach the continuum limit while making the scale of unitarity violation for the deconstructed theory arbitrarily high.\footnote{This contradicts the fact that the five-dimensional theory has an intrinsic cutoff $1/g_5^2$, but this apparent paradox is resolved once we realize that for large enough $\alpha$, the deconstructed theory goes non-perturbative. This necessitates a careful definition of the continuum limit, discussed in Appendix \ref{app:Perturbativity}.}  Consequently, we want to choose $\mbar \propto \alpha$ so that the ratio $\Lambda/\mbar$ is independent of $\alpha$. Indeed, $m_{\rm avg}$ has the correct scaling to compensate for the scaling of $\Lambda$: 

\be
\label{eq:mave}
m_{\rm avg} = \frac{\sqrt{\Tr M^2}}{Ng_4} = \frac{fg\sqrt{2N}}{Ng_4} = \frac{fg\sqrt{2}}{\sqrt{N}g_4} =  f\sqrt{2} \sim \alpha.
\ee
In contrast, $\mlight/g_4$ does not scale with $\alpha$, since $\mlight$ is the same for both the deconstructed theory and the continuum theory (see \Eq{MassSpecContinuum}).

We thus arrive at our final definition of $\R$ in the deconstructed theory,
\be
\label{eq:FinalDefR}
\R = \frac{\Lambda}{\sqrt{\Tr M^2}/Ng_4}.
\ee
By construction, both the numerator and denominator are proportional to $f$, so $\R$ is independent of the (local) deconstructed parameters $f$, $g$, and $N$.\footnote{Our parametric analysis of $\R$ can also be applied to the continuum five-dimensional theory; see \App{app:Parametrics}.}  As desired, $\R$ is directly sensitive to nonlocality mainly through the change in $\Lambda$.

\section{Scattering in theory space}
\label{sec:TMatrixLocal}

In this section, we show how to determine the cutoff scale, defined as the scale of tree-level unitarity violation, by studying tree-level $s$-wave scattering amplitudes.  We then compute the ratio $\R$ for the local moose.  In principle, we could perform these calculations directly in the continuum five-dimensional gauge theory, but in practice, it is much easier to use the deconstructed language of theory space and compute scattering amplitudes for ordinary four-dimensional gauge fields.  

We will make one further simplification by considering amplitudes in the Goldstone equivalence limit.  In this limit, longitudinal gauge boson scattering is dominated by the eaten Goldstones, and these Goldstone modes are simply the nonlinear sigma model fields of the moose.  This limit is justified as long as we consider small enough values of $g_4$ such that corrections to the amplitudes of order $m^2_{\rm KK}/E^2$ are small.  The reason for considering this limit is that when we investigate the effect of nonlocality on $\R$, we will find that there is no change to $\Lambda$ to first order in $\epsilon$.  This fact is difficult to see in the full KK theory, but is straightforward to derive for the Goldstones alone.

\subsection{Partial wave unitarity}
\label{sec:PartialWave}

The leading unitarity-violating pieces of massive gauge boson scattering amplitudes arise from the scattering of longitudinal polarization states. In the limit of high-energy scattering, for sufficiently small values of the gauge coupling $g$, the Goldstone equivalence theorem \cite{Cornwall:1974km, Vayonakis:1976vz} allows us to replace longitudinally-polarized gauge bosons by the corresponding (eaten) Goldstone modes when computing amplitudes.

Because the scattering amplitudes depend nontrivially on the gauge indices, computing the scale of unitarity violation for a given scattering channel $\pi^a \pi^b \to \pi^c \pi^d$ does not give the full information about the scale of unitarity violation for the whole collection of Goldstone modes. Instead, we use a coupled-channel analysis \cite{Lee:1977yc} where we compute the whole scattering matrix $\mathcal{T}^{abcd}$ and define the scale of unitarity violation using its largest eigenvalue $\lambda_{\rm max}$.

The conventional normalization of the scattering matrix for partial waves labeled by $J = 0, 1, \dots$ is (following the notation of \Ref{Chang:2003vs}),
\begin{equation}
\label{PartialWave}
a^{(J)}_{\alpha \beta} = \frac{1}{32\pi}\int_{-1}^1 \langle \alpha | \T | \beta \rangle P_J(\cos \theta) \, d  \cos \theta,
\end{equation}
where $\langle \alpha | \T | \beta \rangle$ is the scattering amplitude between properly normalized in and out states $|\alpha \rangle$ and $| \beta \rangle$, and $P_J(\cos \theta)$ is the $J$-th Legendre polynomial. As in earlier applications of this coupled-channel analysis \cite{Lee:1977yc, Chang:2003vs, Cahn:1991xf, SekharChivukula:2006we}, the $J = 0$ $s$-wave piece provides the strictest bounds. Expressing the amplitudes in terms of the Mandelstam variables $s$, $t = -\frac{s}{2}(1 + \cos \theta)$, and $u = -\frac{s}{2}(1 - \cos \theta)$, the integration in \Eq{PartialWave} for $J = 0$ amounts to the replacement $t, u \to -s/2$ in the amplitude $\T$ and simply contributes a factor of 2 from $d \cos \theta$. 

For unitarity to hold, the largest eigenvalue of the $s$-wave partial amplitude must satisfy\footnote{The precise numerical value of the unitarity bound is not important for our analysis.  If we wished, we could use this freedom to suppress the effects of unknown four-derivative terms, whose amplitudes grow as $p^4$.   For example, imposing $|\text{Re}\, (a^{(0)} )| < 1/4$ (corresponding to the scale of ``half-unitarity violation'') would suppress such terms by a factor of $(1/2)^2 = 1/4$. In our analysis, all that matters is the parametric dependence of $\Lambda_{\rm NL}/\Lambda_{\rm local}$ on $\epsilon$, which is unchanged by such manipulations.}  $|\text{Re}\, (a^{(0)} )| < 1/2$, leading to the condition on the amplitude\footnote{Note that $\lambda(s)$ will be real for the tree-level amplitudes we consider in this paper, so $|\lambda(s)|$ is an absolute value and not a complex modulus.}
\begin{equation}
|\lambda(s)| < 8\pi,
\end{equation}
where $\lambda(s)$ is the largest eigenvalue of the $\T$ matrix expressed as a function of the Mandelstam variable $s$ (after the replacement $t, u \to -s/2$).  For the Goldstone equivalence limit we consider in this paper, the eigenvalues will be linear in $s$, so we can define $\lambda' \equiv \partial \lambda/\partial s$.  By solving for $\sqrt{s}$, we obtain the scale of unitarity violation:
\be
\label{SWaveUnitarity}
\Lambda =  \sqrt{\frac{8\pi}{|\lambda_{\rm max}'|}}.
\ee
Since the $\T$ matrix (expressed as a function of $s$) is related to the $s$-wave piece $a^{(0)}$ by numerical factors, we can work directly with $\T$ and get the scale of unitarity violation from \Eq{SWaveUnitarity}.

\subsection{Local moose scattering matrix}
\label{sec:LocalScatteringMatrix}

To compute the gauge boson scattering matrix for the local moose in the high-energy limit, we can derive Feynman rules directly from the $|D_\mu \Sigma_j|^2$ part of the Lagrangian, expressed in terms of the Goldstone fields $\pi^a_j$. Because we work in the high-energy limit, we need not work in the basis of Goldstones which are eaten by the gauge boson mass eigenstates; rather, we work in a basis which makes the global symmetries of the scattering matrix manifest.\footnote{In fact, \Ref{Chivukula:2002ej} computed all amplitudes in the mass eigenstate basis, but found that only a coupled-channel analysis reproduced the correct unitarity behavior.  The dominant channel is precisely the one we find in the Goldstone basis.}

\begin{figure}[t]
\begin{center}
\includegraphics[scale=0.35]{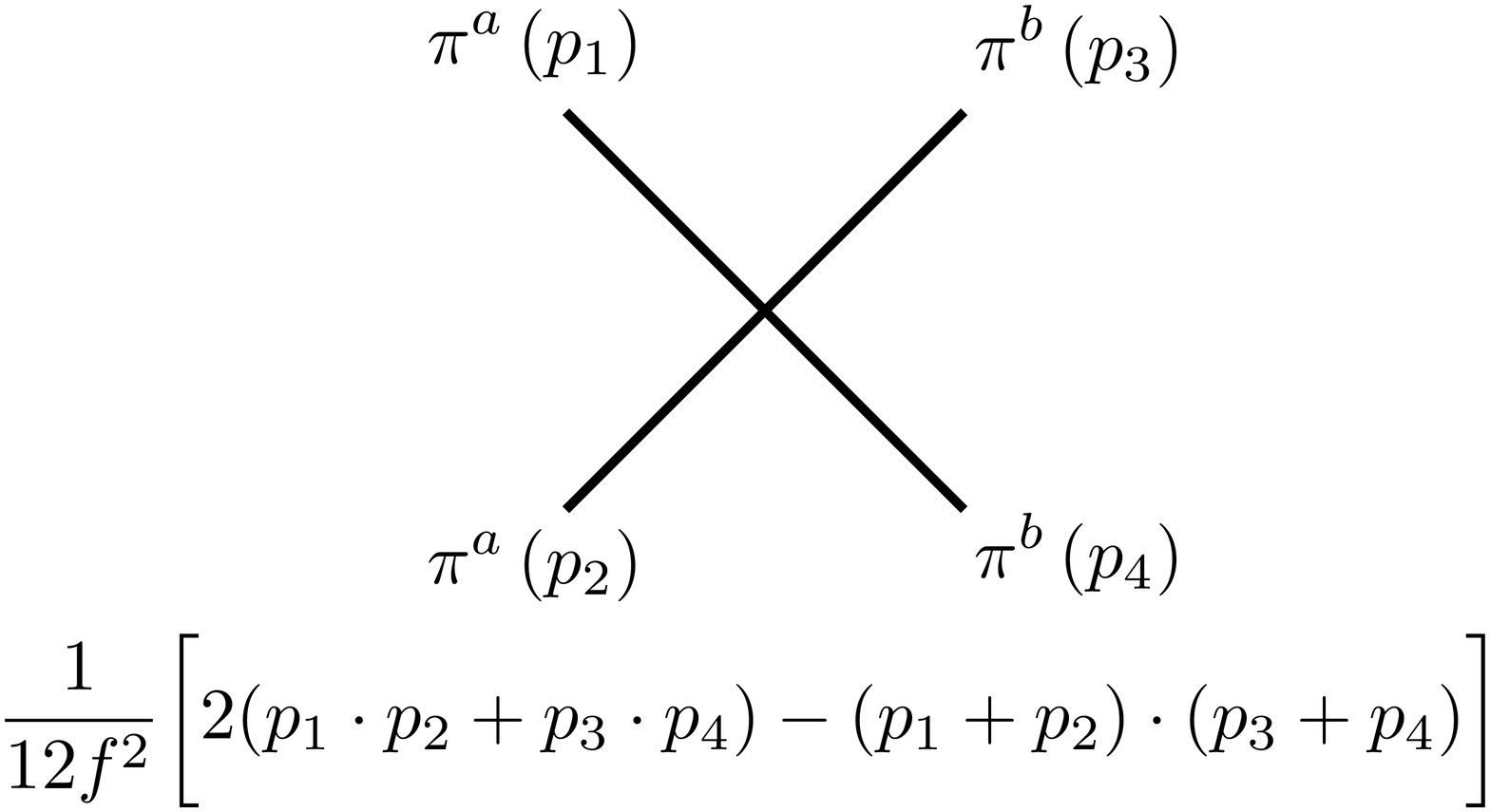}
\caption{Feynman rule for the 4-point interaction contributing to $\SU(2)$ singlet-channel scattering $\pi^a \pi^a \to \pi^b \pi^b$. By convention, all momenta are ingoing.  For the case of $\SU(2)$, the coefficient of the amplitude is independent of the choice of indices so long as $a \neq b$; if $a = b$ the 4-point vertex vanishes identically.}
\label{fig:4ptLocal}
\end{center}
\end{figure}

The $\pi_j \to -\pi_j$ reflection symmetry of the Lagrangian (\ref{CyclicMooseLag}) implies the absence of terms odd in $\pi_j$, and hence the leading-order interaction is a 4-point term,
\begin{equation}
\label{4ptLocal}
 \frac{1}{6f^2}\sum_{j=1}^N\Big [ (\partial_\mu \pi_j^a) \pi_j^b (\partial^\mu \pi_j^c) \pi_j^d - (\partial_\mu \pi_j^a) (\partial^\mu \pi_j^b) \pi_j^c \pi_j^d \Big]\Tr(T^a T^b T^c T^d).
\end{equation}
The Feynman diagram contributing to $\pi \pi \rightarrow \pi \pi$ scattering is shown in \Fig{fig:4ptLocal}.  Notice that for the local Lagrangian, $\pi_i$ and $\pi_j$ are totally decoupled for $i \neq j$.

Now, consider the $\T$ matrix for an SU(2) moose, which will be our starting point for introducing nonlocality in the following section. Because all of the $N$ link fields are decoupled, $\T$ consists of $N$ copies of the $\T$ matrix for one link, allowing us to drop link indices in what follows.  The Goldstone modes are triplets of $\SU(2)$, so the two-particle scattering states decompose into irreducible representations of $\SU(2)$ as $\mathbf{3} \otimes \mathbf{3} = \mathbf{1} \oplus \mathbf{3} \oplus \mathbf{5}$.  Borrowing the language of chiral perturbation theory, the pions $\pi^a$ are in the isospin $I = 1$ representation, and scattering takes place between two-particle states of definite total isospin $I = 0, 1, 2$. 

By Bose symmetry, the $s$-wave scattering amplitude vanishes for the antisymmetric \linebreak $I=1$ states.  The remaining nonzero eigenvalues can be calculated analytically by diagonalizing $\T$:
\be
\lambda_{I=0} = \frac{s}{4f^2}, \qquad \lambda_{I=2} = -\frac{s}{8f^2},
\ee
so the largest eigenvalue (strictly speaking, the eigenvalue of largest magnitude) is in the $I = 0$ sector.  We will refer to the corresponding eigenvector as the \emph{gauge singlet} state\footnote{The additional normalization factor of $1/\sqrt{2}$ accounts for the fact that in the isospin basis, the final state particles for each isospin channel are identical.}
\begin{equation}
\label{SU2Singlet}
|S \rangle  \equiv \sum_{a=1}^3 \frac{1}{\sqrt{3}} \, \frac{1}{\sqrt{2}} |\pi^a \pi^a \rangle.
\end{equation}
As we will see, the fact that the gauge singlet is associated with the largest eigenvalue is the key property which implies the vanishing of the leading-order nonlocal contribution to the scattering matrix. Using $\lambda_{\rm max} \equiv \lambda_{I=0}$ in \Eq{SWaveUnitarity} gives a scale of unitarity violation 
\be
\label{LambdaLocal}
\Lambda_{\rm local}  = 4\sqrt{2\pi} f.
\ee
Note that this is stronger than the naive dimensional analysis bound (\ref{NDAbound}) by a factor\footnote{This leads to a very rough estimate of the effects of four-derivative terms at the cutoff. The NDA bound is the scale at which one-loop effects are comparable to tree-level effects, so at the actual cutoff, one can estimate the size of tree-level four-derivative terms as being suppressed relative to the local two-derivative terms by a factor of $(\Lambda_{\rm local}/\Lambda_{\rm NDA})^2 \approx 1.6$.} of $\sqrt{\pi/2} \approx 1.25$. Furthermore, the largest amplitude for any individual scattering channel $\pi^a \pi^a \to \pi^b \pi^b$ is a factor of 2 smaller than that of the gauge singlet; performing the coupled-channel analysis strengthens the unitarity bounds considerably. Including the results for $\mbar_{\rm local}$ from \Eq{eq:mave}, the normalized ratio $\R$ for the local theory is 

\be
\label{RLocal}
\R_{\rm local} \equiv \frac{\Lambda_{\rm local}}{\mbar_{\rm local} } = \frac{4\sqrt{2\pi} f}{f\sqrt{2}} = 4\sqrt{\pi} \approx 7.09.
\ee

\section{Nonlocality in a cyclic moose}
\label{sec:NSiteNL}

We now turn to the main result of this paper, computing the effects of nonlocality on the ratio $\R$ in a cyclic $N$-site moose.  We focus on the case of gauge group $\SU(2)$ because the group theory objects $\Tr(T^a T^b T^c)$ and $\Tr(T^a T^b T^c T^d)$ which accompany 3- and 4-point Feynman diagrams have no simple expressions except for $\SU(2)$.  Nonetheless, the dependence of $\R$ on the strength of the nonlocal term and the nonlocal length scale is independent of the choice of gauge group, and we comment on the case of general $\SUn$ in \App{app:SUn}.

Consider perturbing the local cyclic Lagrangian (\ref{CyclicMooseLag}) (with equal decay constants $f$) by the nonlocal term discussed in \Sec{sec:NLTerms},
\begin{equation}
\mathcal{O}_{\NL} =\epsilon f^2 \sum_{i=1}^N \mathcal{O}_{N_{\hop}}^{(i)},
\tag{\ref{CyclicONL}}
\end{equation}
where $\mathcal{O}_{N_{\hop}}^{(i)}$ is defined in \Eq{Hopij}. In general, such a perturbation will change the values of $\mbar$ and $\Lambda$ from their values in the local theory; we denote the nonlocal values by $\mbar_{\NL}$ and $\Lambda_{\NL}$. In what follows, we will do a perturbation theory analysis for $\mbar_{\NL}$ and $\Lambda_{\NL}$ to order $\epsilon^2$. We will also only be interested in the leading behavior in $N_{\hop}$, since in the continuum limit, we set $N_{\hop} \equiv N \ell /R$ and take $N \to \infty$.\footnote{Strictly speaking, $N \to \infty$ is not the correct continuum limit, as there is a maximum value of $N$ such that the moose reproduces five-dimensional physics; see \App{app:Perturbativity}.}

The calculations below quickly become technical, so we will summarize the main points.
\begin{itemize}
\item The normalization factor $\mbar$ is not affected by the nonlocal term up to discretization errors, so all dependence on nonlocality is contained in $\Lambda$.   
\item The largest eigenvalue of the scattering matrix \emph{does not} change to first order in $\epsilon$.  This surprising fact is the most important result in this paper, since it implies that $\epsilon = 0$ is a special value.  This result depends only on the particular structure of the gauge singlet state corresponding to the largest unperturbed eigenvalue, and not on the equality of decay constants which we have assumed for convenience. 
\item The leading shift in the largest scattering eigenvalue is always positive and proportional to $\epsilon^2$, and hence the shift in $\Lambda$ goes like $-\epsilon^2$, showing that locality yields a local maximum of $\R$ independent of the sign of $\epsilon$.  
\end{itemize}
For larger values of $\epsilon$ beyond perturbation theory, additional scattering channels become relevant and we do not have a closed form solution for $\Lambda_{\NL}$.  While we cannot rule out the possibility that the global maximum of $\R$ might occur with strong nonlocal terms, our numerical studies suggest that $\epsilon = 0$ is indeed the global maximum of $\R$. 

\subsection{The mass matrix with nonlocality}
\label{sec:NSiteNLMass}

The effect of the nonlocal term \eq{CyclicONL} on the spin-1 mass matrix is straightforward.  Any individual hopping term $\mathcal{O}_{N_{\hop}}^{(i)}$ contains new covariant derivatives $D_\mu \Sigma_i$ and $D^\mu \Sigma_{i + N_{\hop}}$.  After setting $\Sigma_i$ to its vev $\langle \Sigma_i \rangle = \one$, the contribution of $\mathcal{O}_{N_{\hop}}^{(i)}$ to the mass matrix is
\begin{equation}
 \Delta M^2  =  \epsilon g^2
\begin{pmatrix}0 &  &  & \cdots &  &  &  & \cdots & 0  \\ &  &  &  &  & 1 & -1 &  \\ & \vdots &  & \ddots &  & -1 & 1 &  \\ &  & 1 & -1 &  &  &  &  \\ &  & -1 & 1 &  &  &  &  \\ & \vdots  &  &  &  &  &  & \ddots
\\0 &  &  &  &  &  &  & & 0\end{pmatrix},
\end{equation}
where the upper-left corners of the $2 \times 2$ blocks are inserted at the positions $(i, i + N_{\hop})$ and $(i + N_{\hop} ,i)$.  Unless $N_{\hop}$ is equal to 0 or $\pm1$ modulo $N$, the perturbation does not affect the diagonal elements, and hence $\Tr M^2$ is unchanged.  Thus, the scale $\mbar$ is identical to the local case, as long as $1 < N_{\hop} < N-1$:
\be
\mbar_{\rm NL} = \mbar_{\rm local} \qquad (1 < N_{\hop} < N-1).
\ee
As desired from \Sec{sec:Diagnostic}, $\mbar$ is insensitive to the nonlocal perturbation, allowing $\R$ to depend on nonlocality only through $\Lambda$.  The case $N_{\hop} \equiv \pm 1$ (mod $N$) can be considered as a discretization error arising from the fact that an $N$-site moose for finite $N$ has a minimum nonlocal length scale $\ell = a$, where $a$ is the lattice spacing.

\subsection{Perturbation theory for scattering matrix eigenvalues}

In general, numerical methods are needed to compute the exact eigenvalues of the scattering matrix as a function of the nonlocal coefficient $\epsilon$.  Here, we will resort to perturbation theory in $\epsilon$, considering the effect of small nonlocal terms.   By ``perturbation theory,'' we mean finding the largest eigenvalue of the scattering matrix, using techniques familiar from nonrelativistic quantum mechanics.\footnote{Of course, throughout this paper we are using ordinary perturbation theory appropriate to quantum field theory to compute scattering amplitudes from tree-level Feynman diagrams.}   Taking the finite-dimensional $\T$ matrix as our ``Hamiltonian,'' we can calculate the scattering matrix eigenvalues order-by-order in $\epsilon$ in terms of the unperturbed scattering eigenstates from the local Lagrangian.

As we will see below, the nonlocal terms change the kinetic matrix for the $\pi$ fields, requiring field redefinitions to compute scattering between canonically normalized states.  Before the field redefinitions, only the nonlocal terms carry factors of $\epsilon$, but expanding the field redefinition matrix as a series in $\epsilon$, both the local and nonlocal terms can contribute to tree-level scattering amplitudes at \emph{all} orders in $\epsilon$.  To keep track of $\epsilon$ in the total amplitude, we have to apply perturbation theory to the somewhat unusual case of a matrix perturbed by contributions to all orders in $\epsilon$. To second order in $\epsilon$, the scattering matrix $\T$ and its largest eigenvalue $\lambda$ are
\begin{align}
\T & = \T_0 + \epsilon \T_1 + \epsilon^2 \T_2 + \cdots, \\
\lambda & = \lambda^{(0)} + \epsilon \lambda^{(1)} + \epsilon^2 \lambda^{(2)} + \cdots.
\end{align}

The fact that the nonlocal term still preserves the cyclic symmetry determines the structure of the eigensystem of the scattering matrix. Indeed, the eigenstates of the full scattering matrix must respect the cyclic symmetry, and in particular there is no mixing of states with different eigenvalues under the operator which implements translations from site $i$ to $i+1$. Thus we can work in the subspace of cyclically-invariant states with eigenvalue zero under this operator. An orthonormal basis of two-particle $\SU(2)$ singlet states $| q \rangle$ spanning this subspace is
\begin{equation}
\label{CyclicInvtStates}
| q \rangle  \equiv \frac{1}{\sqrt{N}} \sum_{i = 1}^{N} \sum_{a = 1}^{3}\frac{1}{\sqrt{6}} |  \pi_i^a \, \pi_{i + q}^a \rangle, \qquad  q = 0, 1, \dots, N-1.
\end{equation}
For instance, in this notation we have
\be
|0 \rangle = \frac{1}{\sqrt{N}} \sum_{i=1}^{N} |S_i \rangle,
\ee
where $|S_i \rangle$ are the gauge singlet states introduced in \Eq{SU2Singlet}, now with a link index.  In the local theory, all of the $\SU(2)$ singlets $| S_i \rangle$ are degenerate with eigenvalue
\be
\lambda^{(0)} = \frac{s}{4f^2},
\ee
so this is also the eigenvalue of $|0 \rangle$. By the arguments above, the eigenstate of the nonlocal theory with the largest eigenvalue must contain $|0 \rangle$, plus order $\epsilon$ contributions from $| q \rangle$:
\be
\label{eq:ScatteringEigenvector}
|\lambda \rangle = |0 \rangle + \epsilon \sum_{q=1}^{N-1} c_q^{(1)} |q \rangle + \cdots.
\ee

With the correct basis in hand, the lowest-order contribution to $\lambda$ at first order in $\epsilon$ is
\be
\label{eq:order1shift}
\lambda^{(1)} = \langle 0 | \T_1 | 0 \rangle.
\ee
However, as we show in \Sec{sec:NLDiagnostic}, this contribution \emph{vanishes identically} for $N_{\hop} < N$, due to a trace-related cancellation peculiar to the gauge structure of the singlet state and not because of any symmetry argument one can make at the Lagrangian level. Indeed, first-order shifts are generic for the other eigenvalues belonging to the $\bm{3}$ and $\bm{5}$ components of $\T$, and as we will see the vanishing of the shift for the singlet requires a detailed look at the structure of the relevant Feynman rules.  The fact that the singlet state is associated with the largest eigenvalue is a necessary condition for locality to maximize $\Lambda$; otherwise, $\Lambda_{\rm NL}$ could be larger or smaller than $\Lambda_{\rm local}$ depending on the sign of $\epsilon$. 

Because of the first-order cancellation, we must go to second order in $\epsilon$.  The first-order eigenvector coefficients for \Eq{eq:ScatteringEigenvector} are
\begin{equation}
\label{CyclicFirstOrderEigenvec}
c_q^{(1)} = \frac{\langle 0 | \T_1 | q \rangle}{\lambda^{(0)}},
\end{equation}
which is the usual formula from ordinary perturbation theory.  Note that the usual energy denominator $\lambda^{(0)} - \lambda_q^{(0)}$ is replaced by just $\lambda^{(0)}$ because the $|q \rangle$ states for $q \neq 0$ have \emph{zero} eigenvalues in the unperturbed $\T$ matrix.  The second-order eigenvalue shift $ \lambda^{(2)}$ is
\begin{equation}
\label{SecondOrderEigenvalShift}
\lambda^{(2)} = \langle 0 | \T_2 | 0 \rangle + \sum_{q=1}^{N-1} \frac{ |\langle 0 | \T_1 | q \rangle|^2}{\lambda^{(0)}},
\end{equation}
which is the usual formula plus an extra term $\langle 0 | \T_2 | 0 \rangle$, which in this context contains contributions from both the local and nonlocal terms. As we will see below, the contribution from the eigenstates $|q \rangle$ is essential, giving the leading behavior in $N_{\hop}$ and illustrating once again the necessity of doing a coupled-channel analysis rather than focusing on single scattering channels.

\subsection{The scattering matrix with nonlocality}
\label{sec:NSiteNLScattering}

The nonlocal term (\ref{CyclicONL}) has several interesting effects on the scattering matrix and, by extension, on the scale of unitarity violation $\Lambda_{\NL}$:
\begin{itemize}
\item $\mathcal{O}_{\NL}$ breaks the reflection symmetry $\pi_j \to - \pi_j$, so 3-point terms are present, unlike in the local Lagrangian.
\item Despite only transforming under the gauge symmetries corresponding to sites $i$ and $i+N_{\hop}$, the operator $\mathcal{O}^{(i)}_{N_{\hop}}$ contains all the link fields between sites $i$ and  $i+N_{\hop}$, and hence several new intermediate states in 3-point diagrams are available to contribute to the gauge singlet scattering channel. These will cause the matrix elements $\langle 0 | \T_2 | 0 \rangle$ to scale like $N_{\hop}$, though $\langle 0 | \T_1 | q \rangle \sim N_{\hop}$ will turn out to give the dominant contribution to $\lambda^{(2)}$.
\item As mentioned above, since $\mathcal{O}_{\NL}$ contains $\partial_\mu \pi$ terms, it changes the kinetic matrix of the $\pi$ fields.   One must perform a field redefinition in order to recover canonical kinetic terms corresponding to the physical fields which participate in scattering. 
 \end{itemize}
 
 Before the necessary field redefinitions, the operators present in the Lagrangian are as follows, writing $i'$ for $i+ N_{\hop}$ for ease of notation:
 \begin{itemize}
 \item Local 4-point terms, independent of $\epsilon$:
 \end{itemize}
 \be
\mathcal{L} \supset \frac{1}{6f^2}\sum_{i=1}^N \Big[\partial \pi_i^a \pi_i^b \partial \pi_i^c \pi_i^d - \partial \pi_i^a \partial \pi_i^b \pi_i^c \pi_i^d \Big]\Tr(T^a T^b T^c T^d).
\tag{\ref{4ptLocal}}
\ee
\begin{itemize}
\item Nonlocal 3-point terms, proportional to $\epsilon$:
\end{itemize}
\be
\label{3ptNL}
\mathcal{L} \supset  -\frac{i\epsilon}{f}\bigg \{ \sum_{i=1}^N \Big [ \,(\partial \pi_i^a \partial \pi_{i'}^b -\partial \pi_{i'}^a\partial \pi_i^b)\bigg (\pi_i^c + \pi_{i'}^c + 2\sum_{k=i+1}^{i'-1}\pi_k^c \bigg ) \Big ]\bigg \} \Tr(T^a T^b T^c).
\ee
\begin{itemize}
\item Nonlocal 4-point terms, proportional to $\epsilon$:\footnote{We have abbreviated the third line because it does not contribute to any relevant amplitudes in our calculation. The omitted terms simply involve various permutations of the group and site indices.}
\end{itemize}
\begin{align}
\label{4ptNL}
& \mathcal{L} \supset \frac{\epsilon}{f^2} \bigg \{  \sum_{i =1}^N \Big [ \, \frac{2}{3} \partial \pi_i^a \pi_{i'}^b \partial \pi_{i'}^c \pi_{i'}^d + \frac{2}{3} \partial \pi_i^a \pi_{i}^b \partial \pi_{i'}^c \pi_{i}^d - \frac{1}{3} (\partial \pi_i^a \partial \pi_{i'}^b + \partial \pi_{i'}^a \partial \pi_{i}^b) ( \pi_{i}^c\pi_{i}^d + \pi_{i'}^c \pi_{i'}^d) \nonumber  \\
&\hspace{25mm} +  \frac{1}{2} \big ( \partial \pi_i^a \pi_{i'}^b  \partial \pi_{i'}^c \pi_{i}^d + \partial \pi_i^a \pi_i^b \partial \pi_{i'}^c \pi_{i'}^d  - \partial \pi_{i'}^a \partial \pi_i^b \pi_i^c \pi_{i'}^d  - \partial \pi_i^a \partial \pi_{i'}^b  \pi_{i'}^c \pi_{i}^d \big ) \nonumber \\
&\hspace{25mm} + \big(\partial \pi_{i'}^a \pi_{i'}^b  \partial \pi_{i}^c + (\textrm{3 similar}) - \partial \pi_{i}^a \partial \pi_{i'}^b  \pi_{i'}^c - (\textrm{3 similar}) \big)\sum_{k=i+1}^{i'-1}\pi_k^d \nonumber \\
&\hspace{25mm} + \sum_{k=i+1}^{i'-1} \Big (2\, \partial \pi_i^a \pi_k^b \partial \pi_{i'}^c \pi_k^d - (\partial \pi_{i}^a  \partial \pi_{i'}^b + \partial \pi_{i'}^a  \partial \pi_{i}^b)\pi_k^c \pi_k^d \Big ) \Big ]\bigg \} \Tr(T^a T^b T^c T^d).
\end{align}
Lorentz indices are suppressed everywhere since the Lorentz structure of all terms is identical, summation over the group indices $a,b,c,d$ is assumed, and the group theory factors for $\SU(2)$ are ($a,b,c,d = 1,2,3$)
\be
\label{GpTheoryFactors}
\Tr(T^a T^b T^c)   = \frac{i}{4}\epsilon^{abc}, \qquad \Tr(T^a T^b T^c T^d)  = \frac{1}{8} \left (\delta^{ab} \delta^{cd} + \delta^{ad} \delta^{bc} - \delta^{ac} \delta^{bd} \right ).
\ee

\begin{figure}[tc]
\begin{center}
\includegraphics[scale=0.3]{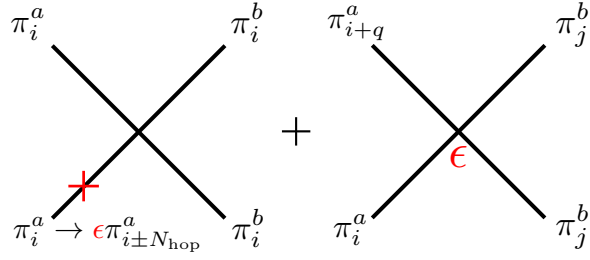}
\caption{Diagrams contributing to the singlet scattering matrix at order $\epsilon$, $\langle 0 | \T_1| q \rangle$. The first term is a local 4-point vertex with a field redefinition insertion on one of the legs, and the second term is a nonlocal 4-point vertex.  These diagrams vanish for $\langle 0 | \T_1| 0 \rangle$, such that there is no order $\epsilon$ contribution to singlet scattering.  For large $N_{\rm hop}$, these diagrams yield the dominant contribution to singlet scattering at order $\epsilon^2$ through terms in \Eq{SecondOrderEigenvalShift} proportional to $|\langle 0 | \T_1| q \rangle|^2$.}
\label{fig:FirstOrder}
\end{center}
\end{figure}

\begin{figure}[tc]
\begin{center}
\includegraphics[scale=0.38]{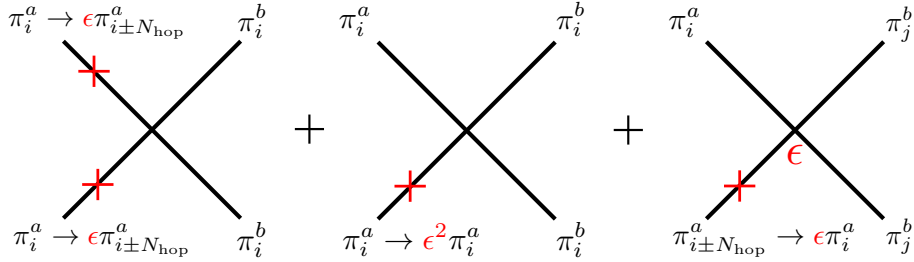}
\caption{4-point vertices contributing to the singlet scattering matrix at order $\epsilon^2$, $\langle 0 | \T_2 | 0 \rangle$. The first term is a local interaction with two order-$\epsilon$ field redefinitions, the second term is a local interaction with one order-$\epsilon^2$ redefinition, and the third term is a nonlocal interaction with one order-$\epsilon$ redefinition. }
\label{fig:SecondOrderFourPoint}
\end{center}
\end{figure}

\begin{figure}[tc]
\begin{center}
\includegraphics[scale=0.35]{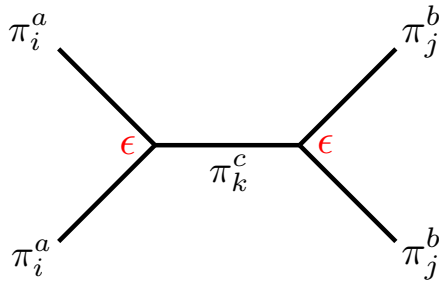}
\caption{Contribution of 3-point (nonlocal) terms to the singlet scattering matrix at order $\epsilon^2$, $\langle 0 | \T_2 | 0 \rangle$.}
\label{fig:SecondOrderThreePoint}
\end{center}
\end{figure}

To take into account the field redefinition, we show in \App{app:Zfactor} that to order $\epsilon^2$, the wavefunction factor $Z^{-1/2}_{ij}$ is
\be
\label{CyclicFieldRedef}
Z^{-1/2}_{ij} = \delta_{ij} - \frac12 \epsilon  \left ( \delta_{i, j+N_{\hop}} + \delta_{i,j-N_{\hop}} \right ) + \frac38 \epsilon^2 \left ( 2\delta_{ij} + \delta_{i, j+2N_{\hop}} + \delta_{i,j-2N_{\hop}} \right ).
\ee
Note in particular that the diagonal elements $Z^{-1/2}_{ii}$ have no order-$\epsilon$ components. After the field redefinitions, we have the following contributions to the gauge singlet piece of the scattering matrix:
\begin{itemize}
\item Order $\epsilon^0$: local 4-point diagrams only, with no field redefinitions (\Fig{fig:4ptLocal}).  These contribute only to $\lambda^{(0)}$.
\item Order $\epsilon$: local 4-point diagrams with order-$\epsilon$ field redefinitions and nonlocal 4-point diagrams with no field redefinitions (\Fig{fig:FirstOrder}).  Since $\langle 0 | \T_1 | 0 \rangle$ vanishes, these only contribute to $\langle 0 | \T_1 | q \rangle$, which dominates $\lambda^{(2)}$.
\item Order $\epsilon^2$: local 4-point diagrams with order-$\epsilon^2$ field redefinitions and nonlocal 4-point diagrams with order-$\epsilon$ field redefinitions (\Fig{fig:SecondOrderFourPoint}), diagrams with two nonlocal 3-point vertices and no field redefinitions (\Fig{fig:SecondOrderThreePoint}).  These contribute to $\langle 0 | \T_2 | 0 \rangle$, which give a subleading (in $N_{\hop}$) correction to $\lambda^{(2)}$.  Note that the first appearance of 3-point terms comes at order $\epsilon^2$.
\end{itemize}

\subsection{Leading effects on the normalized cutoff}
\label{sec:NLDiagnostic}

We now have all of the ingredients to calculate $\lambda$ to order $\epsilon^2$.  First, we can show that the first-order shift vanishes for the singlet state as long as $N_{\hop} < N$.  As mentioned above, we must consider the local terms with field redefinitions and the nonlocal terms separately. The local terms (\ref{4ptLocal}) are all even in $\pi_i$, and so after field redefinitions according to (\ref{CyclicFieldRedef}), the order-$\epsilon$ pieces are odd in $\pi_i$, and thus do not contribute to the matrix element $\langle 0 | \T_1 | 0 \rangle$ between singlet states. The relevant nonlocal terms  in \Eq{4ptNL} would be of the form $\epsilon \partial \pi_i^a \pi_i^a \partial \pi_{i\pm N_{\hop}}^b \pi_{i\pm N_{\hop}}^b$ for fixed group indices $a$ and $b$, but these terms vanish identically due to cyclicity of the trace.\footnote{This argument breaks down when $N_{\hop} \geq N$, because $\mathcal{O}_{\NL}$ contains the redundant kinetic terms discussed in \Sec{sec:NLTerms}, which add extra terms to the second line of \Eq{4ptNL} and destroy the trace-related cancellation. This gives a contribution to the scattering matrix which goes like $\epsilon k^2$, where $k$ is the ``winding number'' around the fifth dimension. The structure of the field redefinition matrix also changes for $N_{\hop} = kN$, giving a further local contribution.}  This can be seen explicitly by setting the appropriate indices equal in the second line of \Eq{4ptNL} and using $\Tr(T^a (T^b)^2 T^a) = \Tr((T^a)^2 (T^b)^2)$.  So there are no contributions to order $\epsilon$ in the singlet channel:
\be
\lambda^{(1)} = 0.
\ee
This result depends only on the cyclicity of the trace and not on the particular form of the group theory factors \eq{GpTheoryFactors}, and holds for any gauge group such that the largest eigenvalue corresponds to the singlet state.\footnote{This is the case for $\SUn$ and is expected to be true for other gauge groups \cite{Chang:2003vs}.} In fact, the vanishing of the first-order singlet amplitude does not depend on the cyclic symmetry.  The relevant nonlocal terms in \Eq{4ptNL} vanish for each individual hopping term $\mathcal{O}_{N_{\hop}}^{(i)}$, and the field redefinitions do not contain any order-$\epsilon$ diagonal terms, as we show in Appendix \ref{app:Zfactor}, thus ensuring that the local terms do not contribute either.

Calculating the second-order eigenvalue shift requires the matrix elements $\langle 0 | \T_1 | q \rangle$ and $\langle 0 |\T_2 | 0 \rangle$ in \Eq{SecondOrderEigenvalShift}. The matrix elements $\langle 0 | \T_1 | q \rangle$ come from terms in the Lagrangian schematically of the form
\be
\epsilon (\pi_j)^2 \partial \pi_i \partial \pi_{i + q}
\ee
for some $i$ and $j$. Since all derivatives in $\mathcal{O}_{\NL}$ are separated by $N_{\hop}$ links, and first-order field redefinitions only shift local link indices by $\pm N_{\hop}$, the only nonzero matrix elements are $\langle 0 | \T_1 | \pm N_{\hop} \rangle$.  Thus, only two terms contribute in the sum over $|q\rangle$ in  \Eq{SecondOrderEigenvalShift}.
We outline the calculation of   $\langle 0 | \T_1 | \pm N_{\hop} \rangle$ in \App{app:FirstOrderEigenvector} for $N_{\rm hop} < N$, which gives\footnote{Note that if $N_{\hop} = N/2$, then $N_{\hop} \equiv -N_{\hop} \ (\textrm{mod} \, N)$, so both Kronecker deltas contribute to the single nonzero coefficient $c_{N_{\hop}}^{(1)}$ and we get an extra factor of 2. This changes some of the formulas below when $N_{\hop} = N/2$, but this special case is most likely an artifact of the latticization.}
\begin{equation}
\label{eq:FirstOrderEigenvec}
c_{q}^{(1)} \equiv \frac{\langle 0  | \T_1 | q \rangle}{\lambda^{(0)}} = \epsilon (2N_{\hop} -1) (\delta_{q, N_{\hop}} + \delta_{q, -N_{\hop}}).
\end{equation}
 As might have been anticipated, introducing the nonlocal length scale $N_{\hop}$ causes the eigenvector for the largest scattering eigenvalue to develop contributions sensitive to this length scale $N_{\hop}$.  The coefficient $c^{(1)}$ depends linearly on $N_{\hop}$ because of the sum over $k$ in \Eq{4ptNL}, which contains $N_{\hop}-1$ link fields between the endpoints of each nonlocal term.

From \Eq{SecondOrderEigenvalShift}, the change in the singlet scattering eigenvalue goes like  $(c^{(1)})^2 \sim \epsilon^2 N_{\hop}^2$, so the leading behavior is quadratic in $N_{\hop}$:
\be
\lambda^{(2)} = \lambda^{(0)} (8\epsilon^2 N_{\hop}^2 + \mathcal{O}(N_{\hop})).
\ee
In particular, this contribution $8\epsilon^2 N_{\hop}^2$ is manifestly positive, since $\lambda^{(0)} = s/4f^2$ is also positive.  Thus, to leading order in $N_{\hop}$, the scattering eigenvalue will always increase, shifting the scale of unitarity violation down.

The matrix element $\langle 0 |\T_2 | 0 \rangle$ is subleading in $N_{\hop}$, so is not relevant for maximizing $\R$ in the continuum limit.  As calculated explicitly in \App{app:SecondOrder}, \Eq{T23pt}, the sum over intermediate states as shown in \Fig{fig:SecondOrderThreePoint} gives a contribution linear in $N_{\hop}$, as do the four-point diagrams from field redefinitions of $\mathcal{O}_{\NL}$ (\Eq{T24pt}), which are essentially the same matrix elements which contribute to $c_{q}^{(1)}$.  Collecting all of these finite $N_\hop$ corrections, plus a local contribution which is independent of $N_{\hop}$ (\Eq{T2Local}) yields
\be
\label{SecondOrderShift}
\lambda^{(2)} = \epsilon^2 \, \lambda^{(0)} \times \left\{ \begin{array}{lr}
8N_{\hop}^2 - 10N_{\hop} + 4, & \quad N_{\hop} < N/2, \\
16N_{\hop}^2 - 26N_{\hop} + 6, & \quad N_{\hop} = N/2,  \\
8N_{\hop}^2 - 10N_{\hop} ,        & \quad N_{\hop} > N/2. 
\end{array}
\right.
\ee
We see that even at finite $N_{\hop}$, $\lambda^{(2)}$ is positive, so the scale of unitary violation uniformly decreases. We have also checked this result numerically by computing the eigenvalues of the full tree-level $\T$ matrix for $N = 3, 4, 5, 6$ and $1 \leq N_{\hop} < N$, with $\epsilon = 10^{-3}$; the numerical studies confirm both the dominance of the singlet channel in the largest eigenvalue, and the analytic results for $c^{(1)}$ and $\lambda^{(2)}$ up to $\mathcal{O}(N_{\hop}^3 \epsilon^3)$ terms.\footnote{These extra terms can be significant even for moderately large values of $N_{\hop}$: for example, for $N=6$ and $N_{\hop} =5$, \Eq{SecondOrderShift} gives $\lambda^{(2)}/\epsilon^2 \lambda^{(0)} = 150$, whereas the full numerical result is $\Delta \lambda /\epsilon^2 \lambda^{(0)} \approx 155.5$.}
As shown in \Sec{sec:NSiteNLMass}, $\mbar$ is unaffected by the nonlocal terms for $|N_{\hop}| > 1$, so the only effect on $\R$ is from $\Lambda$. Writing $N_{\hop} = N \ell /R$ and using \Eq{SWaveUnitarity} to relate $\lambda$ to $\Lambda$, we find the change in $\R$ to leading order in $\epsilon$ and $N$,\footnote{In passing to the continuum in \Eq{eq:RNL}, we ignore the doubling of the coefficient of $N^2$ for $N_{\hop} = N/2$ since this appears to be an artifact of the latticization. We also exclude the cases $\ell = a$ and $ \ell = R-a$ because of the discretization effect in the mass matrix.}
\be
\label{eq:RNL}
\R_{\NL} = \R_{\rm{local}} \times 
\left(1 - 4\epsilon^2 N^2 \frac{\ell^2}{R^2} \right ),  \quad a < \ell < R-a.
\ee
As advertised, maximizing our dimensionless ratio implies locality to lowest order in perturbation theory in $\epsilon$. As argued in \App{app:SUn}, this result persists for arbitrary $\SUn$ up to possible subleading corrections in $N$.

\section{Other uses for the normalized cutoff}
\label{sec:OtherUses}

Moose diagrams have a wide range of phenomenological applications, including low-energy descriptions of QCD \cite{Son:2003et, Chivukula:2004kg,Piai:2004yb}, Higgsless unitarizations of the Standard Model \cite{Csaki:2003dt}, and models of composite Higgs bosons \cite{ArkaniHamed:2001nc, ArkaniHamed:2002pa, ArkaniHamed:2002qy}.  Here, we investigate the implications of our dimensionless ratio $\R$ in two cases of interest:  a three-site description of  $\rho$ mesons in QCD, and deconstructed gauge theories in AdS space.  

\begin{figure}[t]
\begin{center}
\includegraphics[scale=0.45]{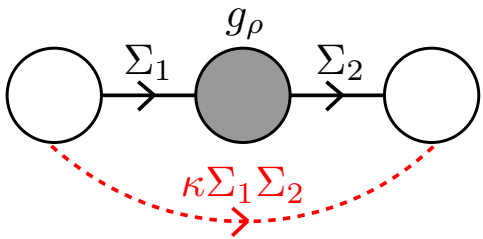}
\caption{The three-site $\SU(2)$ ``hidden local symmetry'' moose describing the pions and $\rho$ mesons of low-energy QCD. The outer two sites are global symmetries, while the middle site is gauged with gauge coupling $g_\rho$. The dashed red line represents the nonlocal term $\Tr|D_\mu \Sigma_1 \Sigma_2|^2$. In contrast to the cyclic moose of \Fig{fig:NLMoose}, we draw this nonlocal term attached to sites, rather than links, to emphasize that it contains redundant kinetic terms.}
\label{fig:3SiteQCD}
\end{center}
\end{figure}

By maximizing $\R$, we can determine the moose parameters which yield an effective theory that stays weakly coupled up to the highest scale.  In the case of three-site QCD, there is an ambiguity in how to choose $\mbar$, but regardless of the choice of $\mbar$, we will see that real-world QCD does not maximize $\R$.  In the case of AdS space, we give preliminary evidence that locality maximizes the cutoff in AdS space, and furthermore that ``$f$-flat'' deconstructions maximize $\R$ among possible local deconstructions.

\subsection{Three-site QCD}
\label{sec:Threesite}

The spin-0 pions and spin-1 $\rho$ mesons of QCD can be modeled using ``hidden local symmetry'' \cite{Bando:1987br}, with a Lagrangian given by
\begin{equation}
\mathcal{L}_{\rm HLS} =  \frac{\tilde{f}^2}{2} \Tr ( |D_\mu \Sigma_1|^2 + |D_\mu \Sigma_2|^2) + \kappa \frac{\tilde{f}^2}{4} \Tr |D_\mu (\Sigma_1 \Sigma_2)|^2 - \frac{1}{2} \Tr (F_{\mu \nu}F^{\mu \nu}).
\end{equation}
This corresponds to the three-site moose shown in \Fig{fig:3SiteQCD}, where the outer sites are global $\SUn_L \times \SUn_R$ chiral symmetries and the middle (shaded) site is gauged with gauge coupling $g_\rho$.  The gauge field $A_\mu$ represents the $\rho$ meson whose longitudinal component is provided by the nonlinear sigma model fields.   The limit $\kappa \to 0$ corresponds to Georgi's vector limit \cite{Georgi:1989xy} where there is an enhanced $\SUn^4$ symmetry when $g_\rho \rightarrow 0$.  The case $\kappa = -1/2$ best approximates the measured masses and interactions of QCD \cite{Brown:1994cq} according to the KSRF relation\footnote{As we have done throughout this paper, we are considering the Goldstone equivalence limit where the gauge boson masses are far below the cutoff, which corresponds to small $g_\rho$. \Ref{Falkowski:2011ua} considers a similar three-site model, but a direct comparison with their analysis is not possible because they consider large $g_\rho$, where $m_\rho$ is of the same order as the cutoff and kinematic thresholds become important.} $m_\rho^2 = 2f_\pi^2 g^2_{\rho \pi \pi}$ \cite{Kawarabayashi:1966kd, Riazuddin:1966sw}. We now investigate whether either limit results from maximizing $\R$. As we intend this example to be only an illustration of our method, and not a sharp statement about the phenomenology of QCD, we restrict to two-derivative terms as we have in the rest of this paper.

As shown in \Eq{eq:choiceOfNonLocal}, we can expand out the $\kappa$ term to express it as kinetic terms for $\Sigma_1$ and $\Sigma_2$ as well as a nonlocal term with $N_{\hop} = 1$. This gives the link field portion of the Lagrangian as
\begin{equation}
\mathcal{L}_{\rm HLS} \supset  \frac{\tilde{f}^2}{4}(2+ \kappa) \Tr ( |D_\mu \Sigma_1|^2 + |D_\mu \Sigma_2|^2) + \kappa \frac{\tilde{f}^2}{4} \widetilde{\mathcal{O}}_{\NL},
\end{equation}
where the nonlocal term is (using our previous notation for the cyclic moose)
\be
\label{NL3site}
\widetilde{\mathcal{O}}_{\NL} \equiv \mathcal{O}_{1}^{(1)} =  2\, \Tr((D_\mu \Sigma_1) \Sigma_2 (D^\mu \Sigma_2^\dagger) \Sigma_1^\dagger).
\ee
Indeed, the three-site case is interesting purely from the point of view of nonlocality, since three is the minimum number of sites needed to have a non-trivial nonlocal interaction. Matching coefficients with our conventions from \Sec{sec:Notation},
\be
\label{Lag3site}
\mathcal{L}_{\rm 3-site} = f^2 ( \Tr|D_\mu \Sigma_1|^2 + \Tr|D_\mu \Sigma_2|^2) + \epsilon f^2 \widetilde{\mathcal{O}}_{\NL},
\ee
gives
\be
f = \frac{\tilde{f}}{2}\sqrt{2+\kappa}, \qquad \epsilon = \frac{\kappa}{2+\kappa},
\ee
 and the KSRF value $\kappa = -1/2$ corresponds to $\epsilon = -1/3$.

To get a dimensionless ratio $\R$, there are two natural choices for mass normalization:  $\mbar = m_\rho/g_\rho$ or $\mbar = f_\pi$. The existence of two natural scales is peculiar to the case of a moose with only one gauged site. There is only one massive gauge boson with mass $m_\rho = \tilde{f} g_\rho$, and one uneaten pion with decay constant $f_\pi =\tilde{f} \sqrt{1+\kappa}$ \cite{Georgi:1989xy}. Since we focus on just the three-site case, there is no continuum scaling argument to decide between $\mbar = m_\rho/g_\rho$ or $\mbar = f_\pi$.

\begin{figure}[tc]
\begin{center}
  \includegraphics[scale = 0.9]{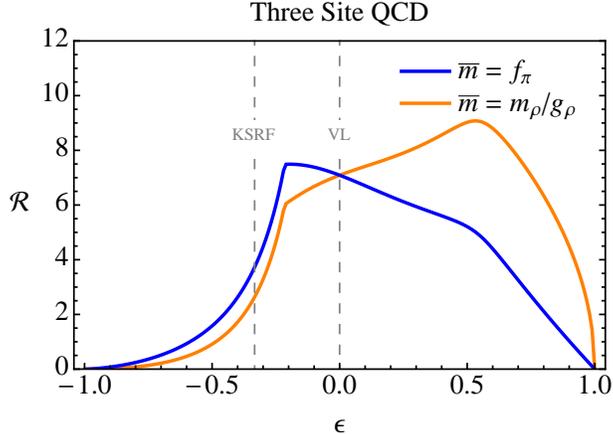}
  \caption{Diagnostic ratio $\R = \Lambda/\mbar$ for the three-site $\SU(2)$ QCD moose as a function of nonlocal coefficient $\epsilon$, comparing two different normalizations. We consider only the range $|\epsilon| < 1$ because the kinetic matrix for the pions becomes degenerate at $\epsilon = \pm 1$.  Also shown are the values of $\epsilon$ corresponding to the vector limit (VL) and the KSRF relation.}
  \label{fig:QCDMoose}
  \end{center}
\end{figure}

We now want to see how $\R$ depends on the nonlocal parameter $\epsilon$.  For the gauge group $\SU(2)$, we can compute the largest eigenvalue of the scattering matrix numerically as a function of $\epsilon$. We find that just as in the case of the cyclic moose described in \Sec{sec:NSiteNL}, the cutoff scale $\Lambda$ is maximized at $\epsilon = 0$, showing the same cancellation in the singlet channel. However, as we have emphasized throughout this paper, $\Lambda$ by itself is not meaningful unless it is normalized. Normalizing to both $f_\pi$ and $m_\rho/g_\rho$, which in terms of our parameters $f$ and $\epsilon$ are 
\be 
f_\pi = f\sqrt{2(1+\epsilon)}, \qquad m_\rho/g_\rho = f\sqrt{2(1-\epsilon)},
\ee
we find the behavior shown in \Fig{fig:QCDMoose}. Because both available normalization scales depend on $\epsilon$ to first order, $\R$ is no longer maximized at $\epsilon = 0$.\footnote{The choice $\mbar = \sqrt{m_\rho f_\pi/g_\rho}$ would give a normalization independent of $\epsilon$ to first order, although this seems somewhat artificial.} The sharp kink at $\epsilon \approx -0.22$ is due to the presence of another scattering channel which begins to dominate the scattering matrix.  Since the nonlocal term preserves the symmetry between links 1 and 2, there are two linearly independent singlet states, 
\begin{equation}
|S \rangle \equiv \frac{1}{\sqrt{2}}(|S_1 \rangle + |S_2 \rangle), \qquad |\widetilde{S}\rangle \equiv \frac{1}{\sqrt{12}} \sum_{a=1}^3 (|\pi_1^a \pi_2^a \rangle + |\pi_2^a \pi_1^a \rangle),
\end{equation}
which are the linear moose analogues of the $|q \rangle$ states. For all $\epsilon$, one linear combination of $|S \rangle$ and $|\widetilde{S}\rangle$ has a positive scattering eigenvalue, and the orthogonal combination has a negative eigenvalue, and their magnitudes are equal at the kink.

For the $m_\rho/g_\rho$ normalization, $\R$ is maximized at $\epsilon \approx +0.53 $, while for the $f_\pi$ normalization, $\R$ is maximized for $\epsilon \approx -0.22$. We find it interesting that the value of $\epsilon$ preferred by the $f_\pi$ normalization has the same sign as the KSRF value, although its magnitude is somewhat smaller, and furthermore that neither normalization prefers the vector limit. We conclude that low-energy QCD does not yield the most weakly coupled two-derivative Lagrangian possible.

\subsection{Warped deconstructions}
\label{sec:Warped}

In \Sec{sec:NSiteNL}, we showed that locality was a special property of cyclic moose models, corresponding to deconstructions of a flat compact dimension.  With an eye toward the AdS/CFT correspondence, we are interested in knowing whether locality is also a special property of warped spaces.  Since the large-$n$ limit of a four-dimensional CFT corresponds to a weakly-coupled AdS$_5$ theory, it is at least plausible that locality in AdS$_5$ might yield the \emph{most} weakly coupled theory possible. In this section, we present circumstantial evidence towards that conclusion, and show that ``$f$-flat'' deconstructions maximize $\R$ in the local case.  We leave a full study of nonlocality in AdS for future work.

To apply our normalized cutoff $\R$ to the deconstruction of nontrivial geometries, we must relax the restriction that the deconstructed gauge couplings and decay constants be equal for every site and link on the moose.  We instead only require that the deconstruction reproduces the AdS$_5$ geometry in the continuum limit.  By maximizing $\R$ over all choices of $f_i$ and $g_i$ which reproduce AdS, we can determine which deconstruction of AdS space stays weakly coupled as long as possible.

Our setup is the Randall-Sundrum (RS1) scenario \cite{Randall:1999ee}, which is an orbifold compactification of AdS$_5$ on $S^1 / Z_2$. Denoting the $(3+1)$-dimensional coordinates by $x$ and the compactified coordinate by $y$ as usual, the metric is
\be
ds^2 = e^{-2ky}dx^2 - dy^2
\ee
with $k$ a constant.  With this metric, the five-dimensional Yang-Mills action is
\be
\label{5dYMRS}
S =  \int d^4 x \,\int_0^{R} dy \, \frac{1}{G^2} \left ( -\frac{1}{2} \Tr ( \Fhat_{\mu \nu}\Fhat^{\mu \nu}) + e^{-2ky}\, \Tr(\Fhat_{\mu 5}\Fhat^{\mu 5}) \right ) ,
\ee
where (unlike in the rest of this paper) we have chosen a normalization for the gauge fields such that the gauge coupling appears outside the kinetic term. In general, the five-dimensional gauge coupling $G$ can vary as a function of the bulk coordinate $y$, corresponding to a nontrivial dilaton profile, but in what follows we will restrict $G$ to be a constant.

\begin{figure}[t]
\begin{center}
\includegraphics[scale=0.45]{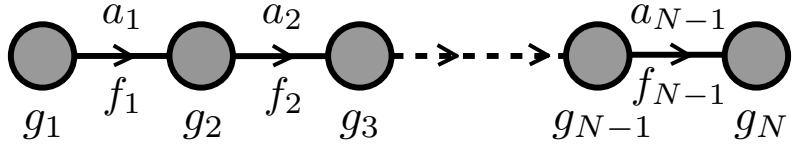}
\caption{Linear $N$-site moose, showing $N$ gauge couplings $g_i$ defined on the sites, but $N-1$ decay constants $f_i$ and lattice spacings $a_i$ defined on links.  By varying the gauge couplings and decay constants, one can describe the deconstruction of warped geometries such as AdS$_5$.}
\label{fig:LinearMoose}
\end{center}
\end{figure}

Choosing a similar normalization for the deconstructed gauge fields, we can write the action for the appropriate $N$-site \emph{linear moose} as
\be
\label{eq:AdSMoose}
S = \int d^4 x \left ( -\frac{1}{2} \sum_{i=1}^N \frac{1}{g_i^2} \Tr F_i^2 + \sum_{i=1}^{N-1}f_i^2 \, \Tr |D_\mu \Sigma_i|^2 \right ),
\ee
where we have allowed the deconstructed gauge couplings $g_i$ to vary.  In contrast to the cyclic moose, this linear moose in \Fig{fig:LinearMoose} has $N$ sites but only $N-1$ links, appropriate for a latticization of an orbifold compactification with boundaries. 

In order for the moose action \eq{eq:AdSMoose} to match the continuum action \eq{5dYMRS}, we have to address some subtleties.  The gauge couplings $g_i$ live on lattice \emph{sites}, whereas the decay constants $f_i$ live on \emph{links}, so there is an ambiguity as to how to determine $g_i$ and $f_i$ in terms of the lattice spacings $a_i$.  We will adopt a convention which preserves the desired relation
\be
\label{4dvs5d}
\frac{1}{g_4^2} = \frac{R}{G^2}
\ee
between the four- and five-dimensional couplings, for all values of the number of sites $N$, though we emphasize that this is but one of many possible choices.  The position along the fifth dimension is given by the lattice spacings $a_i$ ($i = 1$ to $N-1$) through
\be
\label{eq:AdSradiusConstraint}
y_i = \sum_{k=1}^{i}a_k, \qquad R = \sum_{k=1}^{N-1} a_k,
\ee  
and the moose couplings are
\be
\label{eq:AdSgandf}
\frac{1}{g_i^2} = \frac{a_{i-1} + a_{i}}{2 G^2}, \qquad f_i = \frac{\int_{y_{i-1}}^{y_i} e^{-ky}\,dy}{G\sqrt{a_i}},
\ee
where for $g_1$ and $g_N$, we define $a_0 = a_N = 0$, corresponding to the absence of boundary kinetic terms.  Note that the values of $g_i$ and $f_i$ are now determined uniquely in terms of $a_i$.

We can now speculate about locality in AdS$_5$ based on our analysis of \Sec{sec:NSiteNL}.  Since the gauge coupling does not appear in the Goldstone scattering amplitudes in the Goldstone equivalence limit, the cutoff only depends on the decay constants $f_i$ and the coefficients of the nonlocal operators $\mathcal{O}_{N_{\hop}}^{(i)}$.  As we have emphasized, the arguments which give rise to the absence of an $\mathcal{O}(\epsilon)$ correction to $\Lambda$ depend only on the gauge singlet state dominating the scattering matrix, and still hold when the $f_i$ are not equal and there is no cyclic sum of nonlocal terms. This fact alone would be enough to make locality maximize $\R$ in AdS space, as long as there exists a normalization $\mbar$ appropriate to the AdS geometry\footnote{Note that the $\alpha$-scaling argument given in \Sec{sec:Diagnostic} ($f_i \to \alpha f_i$, $g_i \to \alpha g_i$) does not work in an AdS geometry, so a different definition of $\mbar$ is needed.}  which does not receive an $\mathcal{O}(\epsilon)$ correction.

Having tentatively concluded that locality is still a local maximum of $\R$ in AdS$_5$, we can still ask whether different \emph{local} deconstructions (that is, choices of $g_i$ and $f_i$) give higher or lower values of $\R$. One convenient choice is to take equal lattice spacings $a_i \equiv a = R/(N-1)$, which fixes $g_i \equiv g = G/\sqrt{a}$ for all $i$ (except $g_1 = g_N = g\sqrt{2}$).  This ``equal-$a$'' deconstruction can reproduce the physics of RS1 with equal lattice spacings and (essentially) equal gauge couplings, but with decay constants varying like the warp factor $f_j \simeq e^{-kaj}/ag$.  This is in fact the standard latticization used in \Refs{Randall:2002qr,Falkowski:2002cm}.  An alternative possibility would be to take all of the decay constants equal, $f_i \equiv f$, which imposes relations on the $a_i$ and hence on the $g_i$.  Such an ``$f$-flat'' deconstruction was first noted in \Ref{Abe:2002rj}, and investigated in detail in \Ref{SekharChivukula:2006we}; it can reproduce the physics of RS1 as well, at the expense of non-uniform gauge couplings and lattice spacings.

\begin{figure}
  \centering
\includegraphics[width=0.46\textwidth]{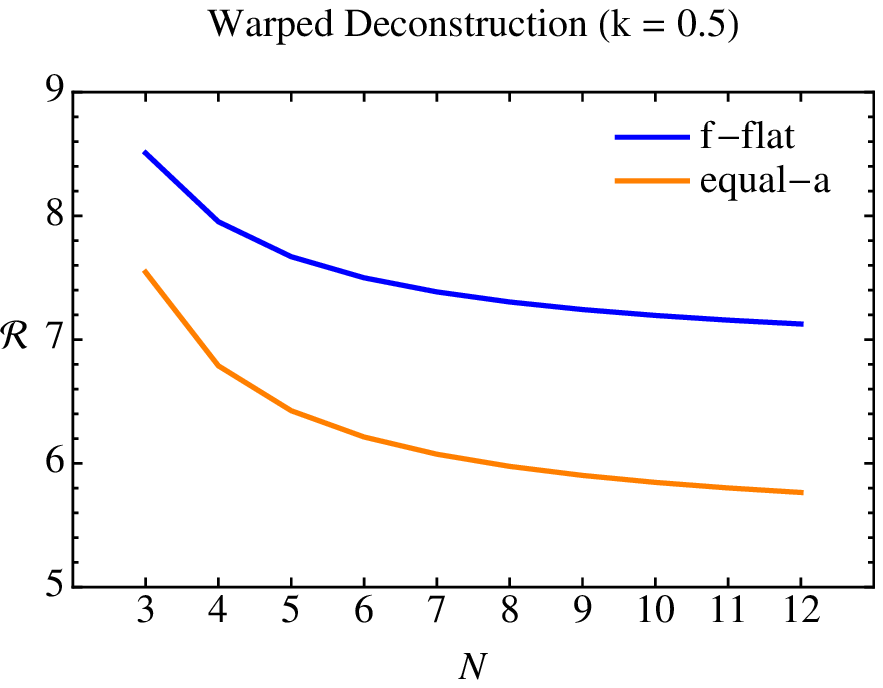} 
  \qquad 
\includegraphics[width=0.47\textwidth]{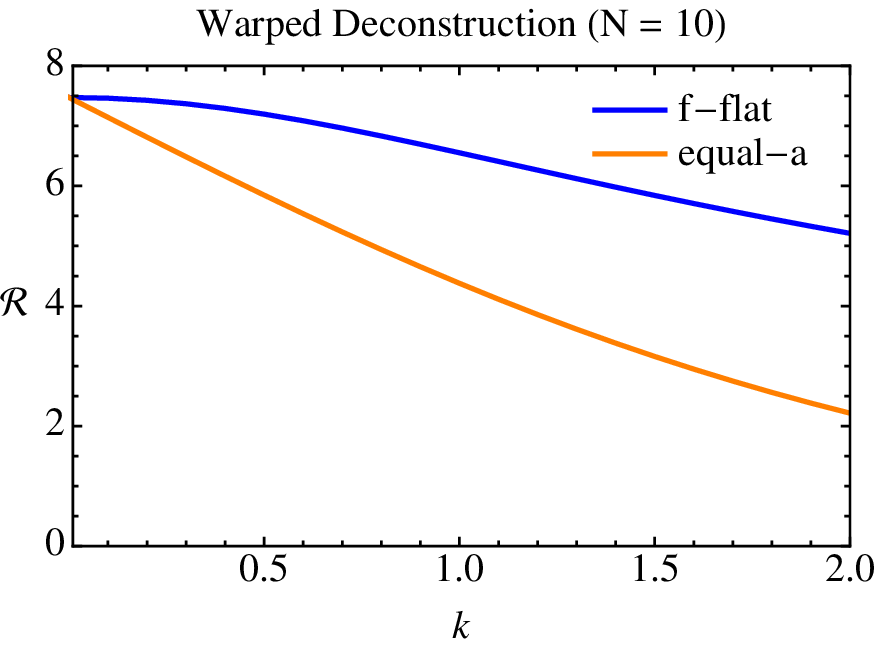}
\caption{Diagnostic ratio $\R$ for two possible deconstructions, $f$-flat and equal-$a$, as a function of number of sites $N$ with $kR = 0.5$ (left) and as a function of $k$ with $N = 10$ (right). The $f$-flat deconstruction is always a maximum of $\R$, and in particular yields a more weakly-coupled theory than the equal-$a$ deconstruction in both cases.}
\label{fig:fvsa}
\end{figure}

Using the value of $\Lambda$ calculated in \Eq{LambdaLocal}, we have
\be
\R^2 = \frac{ g_4^2 \Lambda^2}{\Tr M^2/N^2} = \frac{ 32\pi g_4^2 N^2 \, \text{min} \{ f_i^2 \}}{f_1^2(g_1^2 + g_2^2) + f_2^2(g_2^2 + g_3^2) + \cdots + f_{N-1}^2(g_{N-1}^2 + g_N^2) }.
\ee
We seek to maximize $\R$ (as a function of the $a_i$) leaving the continuum parameters $R$, $g_4$, and $G$ fixed, which means we have to impose the nonlinear constraints in \Eqs{eq:AdSradiusConstraint}{eq:AdSgandf}.  Performing the numerical optimization, we find that the $f$-flat deconstruction is always a maximum of $\R$, up to moderately large values of the AdS curvature, $kR \lesssim 10$. In \Fig{fig:fvsa} (left), we show the optimized value of $\R$ (corresponding to an $f$-flat deconstruction), along with the corresponding value for an equal-$a$ deconstruction for comparison, for $kR = 0.5$ and $g_4 = 1$.\footnote{In this figure, $\R$ appears to be a monotonically decreasing function of $N$, which seems to imply that coarser latticizations are preferred. However, this may be because the normalization which was appropriate for the flat deconstruction is no longer appropriate for the warped deconstruction.} \Fig{fig:fvsa} (right) shows the values of $\R$ for the same $f$-flat and equal-$a$ deconstructions for a fixed number of lattice sites $N = 10$, as a function of $k$ with $R = 1$. As expected, the difference between the two deconstructions goes to zero as $k \to 0$, since in that limit we recover a flat extra dimension where $f$ and $a$ are both equal across the moose.\footnote{Note that the corresponding value of $\R$ is different from that of the cyclic moose, \Eq{RLocal}, because the mass spectrum is different for a linear moose.}

Indeed, from a physical perspective, an $f$-flat deconstruction makes sense. A strong hierarchy of decay constants would mean that some link fields would violate unitarity at a scale far below or above their neighbors, such that the deconstructed description would break down at different scales depending on the bulk coordinate $y$. Since the gauge boson wave functions are supported throughout the bulk, this description is clearly inferior to one where the physics remains weakly coupled everywhere in the bulk up to the same scale.

\section{Conclusion}
\label{sec:Conclusion}

In this paper, we have investigated the effect of nonlocal terms in theory space.  We have shown that such terms, considered as small perturbations, uniformly shift the scale of tree-level unitarity violation $\Lambda$ to lower values compared to the local theory.  Thus, maximizing the normalized cutoff scale $\R \equiv \Lambda/\mbar$ implies a theory which is local in the compact fifth dimension. We have shown this rigorously only in the context of perturbation theory in the strength of the nonlocal terms, but we have also found the same behavior in numerical studies, including the behavior of $\Lambda$ (though not $\mbar$) in the three-site QCD moose. From a four-dimensional perspective, this is a surprising result, since it relies on the absence of an order $\epsilon$ shift in $\Lambda$.  From a five-dimensional perspective, locality in theory space is expected in various UV completions, so it is intriguing that locality is also a special limit of theory space from purely IR considerations. 

We stress that all our results in this paper have been at tree-level. The obvious next step is to analyze the theory at one-loop and examine the behavior of nonlocal terms under renormalization group (RG) flow.\footnote{We thank Allan Adams for suggesting this perspective.}  As we have emphasized, the deconstructed theory is a nonrenormalizable theory  where all interaction terms are irrelevant operators.  If, however, the nonlocal operators disappeared faster in the IR than the local ones, that would give strong support to the statement that locality is a special property of theory space beyond our tree-level analysis.  The fact that the nonlocal operators shift $\Lambda$ downwards suggest that they become stronger than the local terms in the UV, and thus weaker in the IR, so this behavior is at least plausible.\footnote{Such an RG analysis may be difficult because there is not a great deal of room for RG flow between the cutoff and the lowest lying KK modes. For example, by taking the group rank $n$ to be large, the cutoff of the local theory can be made arbitrarily small.}  Indeed, this RG analysis would give a truly IR perspective on locality, as compared to our tree-level analysis, which combines IR physics with the presence of a cutoff towards the UV.\footnote{A one-loop analysis would also provide a definitive answer on the effects of four-derivative terms, both local and nonlocal.} We leave a detailed analysis of nonlocal RG flow to future work.

The effect of dimensionality on locality is also worth exploring.  Imagine deconstructing a four-dimensional theory using a cyclic three-dimensional moose. Yang-Mills in four dimensions has no tree-level cutoff, but the deconstructed nonlinear sigma model has a cutoff related to the inverse lattice spacing $a^{-1}$.  One would want to check that the cutoff is lowered toward the inverse nonlocal length scale $\ell^{-1}$ in the presence of nonlocal terms.  Going one dimension lower, both the local and nonlocal terms become marginal in a two-dimensional moose, and it would be interesting to see whether local operators were marginally relevant while nonlocal ones were marginally irrelevant in the RG perspective described above. Finally, in dimensions higher than four, there are multiple energy scales in play---the nonlinear sigma model cutoff and the intrinsic Yang-Mills cutoff in five or more dimensions---and it would be interesting to investigate the interplay of these scales.

Finally, our analysis might be relevant for understanding the emergence of bulk locality in the AdS/CFT correspondence.  In the standard way of thinking about AdS/CFT, bulk locality is a consequence of the large-$n$ limit of the boundary CFT.  Since the large-$n$ limit also corresponds to the weak coupling limit, it would be quite satisfying if local AdS space were the \emph{most} weakly coupled possibility from the IR point of view.  Our analysis of the various deconstructions of AdS$_5$ provides some compelling hints in that direction.  Because the gauge singlet channel still dominates and maximizing $\R$ leads to $f$-flat deconstructions, we expect that the analysis of nonlocality in AdS space should mirror the analysis of nonlocality in flat space. More rigorously, one should also check that locality is preferred for gravitational modes, in the framework of a deconstructed gravitational dimension \cite{ArkaniHamed:2002sp, ArkaniHamed:2003vb, Randall:2005me, Gallicchio:2005mh}. If the most weakly coupled AdS theories turn out to be the local ones, this would give an IR perspective for why large-$n$ CFTs should have local AdS duals.

\acknowledgments

We thank Allan Adams, Csaba Csaki, Ethan Dyer, and Matthew Schwartz for helpful conversations, and Nima Arkani-Hamed for early inspiration. This work was supported by the U.S. Department of Energy under cooperative research agreement DE-FG02-05ER-41360. J.T. is supported by the DOE under the Early Career research program DE-FG02-11ER-41741. Y.K. is supported by an NSF Graduate Fellowship.

\appendix

\section{Normalized cutoff details}
In this appendix, we explore in more depth some issues regarding the normalized cutoff $\R$.  We first resolve the $\alpha$-scaling issue mentioned in \Sec{sec:Diagnostic} and define the correct continuum limit. We then present a brief analysis of the KK truncation of the continuum five-dimensional theory, and in the process show that $\R$ has a suitable continuum definition. Finally, we consider the effect of imitating an exact KK truncation in the deconstructed picture, and show that in fact a local moose gives a larger value of $\R$.

\subsection{Breakdown of perturbativity}
\label{app:Perturbativity}

As described in \Sec{sec:TMatrixLocal}, the cutoff scale $\Lambda$ is defined in the deconstructed theory by imposing partial-wave unitarity on Goldstone scattering amplitudes in the Goldstone equivalence limit.  However, this limit becomes invalid whenever the scale of scattering becomes comparable to the masses of the gauge bosons, indicating that transverse scattering can no longer be neglected. From \Eq{CyclicMassSpectrum}, the mass of the heaviest deconstructed mode is $2fg$, so Goldstone equivalence holds when
\be
2fg \lesssim 4\pi f  \implies g  \lesssim 2\pi.
\ee
As expected, Goldstone equivalence fails when the gauge coupling gets large and the deconstructed theory becomes non-perturbative.  So while a naive application of the scaling in \Eq{DeconstructedScaling} would seem to indicate that the continuum limit of the deconstructed theory unitarizes the five-dimensional theory up to arbitrarily high scales, in fact the analysis in terms of Goldstone equivalence breaks down for sufficiently large $g$.

As noted in \Ref{Chivukula:2002ej}, a maximum $g$ actually implies that for fixed $g_4$, $N \to \infty$ is \emph{not} the correct continuum limit.  Rather, there is a maximum size of the moose $N$ for which the deconstructed theory resembles the continuum theory,
\be
\label{BoundOnN}
g \lesssim 2\pi  \implies N < \frac{4\pi^2}{g_4^2}.
\ee
Strictly speaking, $N \to \infty$ is only the limit of a free theory with $g_4 = 0$, which can arise from taking the compactification size $R$ to infinity. 

\subsection{KK truncation in the continuum}
\label{app:Parametrics}

Unlike its lower-dimensional counterparts, five-dimensional Yang-Mills is nonrenormalizable because the five-dimensional gauge coupling $g_5$ has negative mass dimension. The cutoff scale $\Lambda_{\rm 5d}$ can be taken as the scale at which gauge boson scattering in the bulk violates tree-level unitarity \cite{SekharChivukula:2001hz}.  For gauge group $\SUn$, we have
\be
\label{5dCutoff}
\Lambda_{\rm 5d} = \frac{96 \pi}{23n}\frac{1}{g_5^2}.
\ee
Apart from numerical factors,\footnote{The depend on the group rank $n$ is similar, though not identical, to the $1/\sqrt{n}$ dependence of the deconstructed theory we find in \Eq{LocalSUnRatio}.} this is expected from dimensional analysis, since $1/g_5^2$ is the only parameter with dimensions of energy in the five-dimensional bulk.  For the purposes of defining a continuum version of the unitarity-violating scale $\Lambda$, we will take $\Lambda_{\rm continuum} \equiv \Lambda_{\rm 5d}$.  Using the deconstruction dictionary (\ref{dictg5}), we find the scaling
\be
\label{LambdaContinuum}
\Lambda_{\rm continuum} =  \frac{48 \pi}{23}\frac{1}{g_5^2} =  \frac{48 \pi}{23} \frac{f}{g},
\ee
where we have taken $n = 2$ to compare with the SU(2) moose considered in the text.

Remarkably, this same unitarity-violating behavior is also manifest in the KK description after compactification.  As shown in \Ref{SekharChivukula:2001hz}, one can define the cutoff of the KK theory by the mass of the heaviest weakly-coupled KK mode, which gives a cutoff $\Lambda_{\rm KK}$ identical to \Eq{5dCutoff} up to numerical factors.  From the four-dimensional KK perspective, it is surprising that the cutoff $\Lambda_{\rm KK}$ can be expressed purely in terms of the combination $g_5 = g_4\sqrt{R}$, but this is necessary for consistency between the various definitions of $\Lambda$.

We now seek a continuum definition of $\R$ which arises from KK truncation. To define the dimensionless ratio $\R$, we also need a continuum definition of $\mbar$, and we can use the modified continuum limit (\ref{BoundOnN}) to suitably define an average mass scale.  In the KK theory, it is impossible to push the KK tower below the cutoff because as mentioned above, the cutoff is \emph{defined} by the mass of one of the KK modes.  Thus, the number of modes we sum over to construct the average mass $\mbar$ should be the number of modes below the cutoff $\Lambda_{\rm continuum}$. Since KK levels are equally spaced by $2\pi/R$, and each level is twofold degenerate, we have
\be
N = 2\,\frac{\Lambda_{\rm continuum}}{2\pi /R} =   \frac{48}{23} \frac{1}{g_4^2}.
\ee
Reassuringly, this is (parametrically) the same continuum limit found in \Eq{BoundOnN} by independent arguments, and in fact is well below the maximum allowed value.  Using \Eq{DeconstructedMbar} and interpreting the trace as a sum over the first $N$ nonzero KK modes (that is, over KK levels $1 \leq k \leq N/2$, with a factor of 2 for degeneracy), we have
\be
\mbar_{\rm continuum} = \frac{2}{Ng_4}\sqrt{\sum_{k=1}^{N/2} \, \left (\frac{2\pi k}{R} \right)^2} = \frac{\pi}{Rg_4^2} \, \sqrt{\frac{32}{23} +\mathcal{O}(g_4^2)},
\ee
and since we are assuming $g_4 \ll 1$ for perturbativity we will drop the $\mathcal{O}(g_4^2)$ terms in the last expression. Using the relation (\ref{4d5d}) between the four- and five-dimensional couplings,
\be
\label{MbarContinuum}
\mbar_{\rm continuum} = \pi \sqrt{\frac{32}{23}} \frac{1}{g_5^2} = \pi \sqrt{\frac{32}{23}} \frac{f}{g}.
\ee

Both $\Lambda_{\rm continuum}$ and $\mbar_{\rm continuum}$ have the same parametric dependence, so the normalized cutoff $\R = \Lambda/\mbar$ is independent of the continuum parameters $R$, $g_4$, and $g_5$, as desired:
\be
\R_{\rm continuum} = \sqrt{\frac{72}{23}} \approx 1.77.
\ee
While the continuum definitions of $\Lambda$ and $\mbar$ both differ from their deconstructed counterparts (\ref{LambdaLocal}) and (\ref{eq:mave}) by a factor of $g$, the ratio $\R$ has the same parametric behavior, and numerically is of the same order of magnitude.

\subsection{KK truncation in deconstruction}
\label{app:KKtruncDeconstruct}

The idea of defining $\Lambda$ through KK truncation raises the question of whether we might obtain a better deconstructed effective theory by requiring the deconstructed mass spectrum to \emph{exactly} match a truncated KK spectrum, rather than having the deconstructed spectrum dictated by locality and only approximately matching the KK spectrum.\footnote{Indeed, in the case of gravitational deconstruction of flat space, a truncated KK spectrum has improved UV behavior compared to strictly local interactions \cite{Schwartz:2003vj}.}   Such an analysis is beyond the perturbative study in \Sec{sec:NSiteNL} since large nonlocal terms are needed to match the spectrum and interactions of the first $N$ modes in the KK tower. Furthermore, the analysis is complicated by the fact that there is an ambiguity in the choice of nonlocal terms. The \linebreak $(N-1)/2$ matching conditions\footnote{Here, we consider only $N$ odd, so that each nonzero mass eigenvalue in the deconstructed theory is twofold degenerate as in the KK truncation.} for the nonzero KK modes are degenerate between $N_{\hop} = k$ and $N_{\hop} = N-k$, and the single condition which breaks this degeneracy is the winding mode 4-point vertex matching, leaving a total degeneracy of $(N-3)/2$ in the nonlocal coefficients.  Because of this degeneracy, one can choose to have all but one of the nonlocal coefficients vanish for $N_{\hop} > N/2$, though this requires adding a kinetic term for the winding mode:
\be
\mathcal{L} \supset c_0 \sum_{i=1}^N |D_\mu (\Sigma_i \Sigma_{i+1} \cdots \Sigma_1 \cdots \Sigma_{i-1})|^2 = c_0 \left ( \sum_{i=1}^N |D_\mu \Sigma_i|^2 + \sum_{k=1}^{N-1} (N-k) \mathcal{O}_{N_\hop = k} \right ).
\ee

In our numerical studies, we have found that $\R$ for the local moose is strictly larger than the corresponding value for a KK truncation.\footnote{We compare values of $\R$, rather than values of the unnormalized cutoff $\Lambda$, because matching the KK spectrum requires the addition of large $N_{\hop} = \pm 1$ terms, which change the trace of the mass matrix. Furthermore, the ambiguity in the definition of $\tilde{f}$ mentioned in \Sec{sec:Diagnostic} cancels in the dimensionless ratio $\R$.}  For example, in a 3-site cyclic SU(2) moose, we can fix both of the nonlocal coefficients using the mass matrix and winding mode matching conditions. We find numerically that $\R^{(3)}_{\rm KK} \approx 5.17$, as compared to $\R_{\rm local} \approx 7.09$ from \Eq{RLocal}.\footnote{The nonlocal coefficients required are $c_1 \approx -0.461$ for $N_{\hop} = 1$ and $c_2 \approx 0.115$ for $N_{\hop} = 2$, which are well outside the perturbative regime. Indeed, the largest eigenvalue of the scattering matrix is twofold degenerate, meaning that the singlet channel no longer dominates.} At least in this example (and in all other examples we have studied numerically), we conclude that KK truncation does not dictate the most weakly-coupled theory, and a local moose yields better behavior.

\section{Scattering matrix results}
\label{app:Scattering}

In this appendix, we collect intermediate results used in the calculation of the nonlocal scattering matrix in \Sec{sec:NSiteNL}.  We will also show a generalization of our result to $\SUn$, showing that maximizing $\R$ implies locality in the large-$N$ limit.  In what follows, we will only calculate $s$-wave contributions as discussed in \Sec{sec:PartialWave}, allowing us to express all amplitudes in terms of the Mandelstam variable $s$, using the replacements $t, u \to -s/2$.

\subsection{Field redefinitions}
\label{app:Zfactor}

The nonlocal perturbation in \Eq{CyclicONL} introduces kinetic mixing between the Goldstone modes at order $\epsilon$.  To put the Lagrangian in canonical form, we require field redefinitions given by the matrix $Z^{-1/2}$, where the kinetic or $Z$-factor matrix is
\begin{equation}
Z_{ij} = \frac{\partial^2 \mathcal{L}}{\partial (\partial_\mu \pi_i) \partial(\partial^\mu \pi_j)}.
\end{equation}
For the local $N$-site moose with the nonlocal perturbation (\ref{CyclicONL}), we have
\begin{equation}
Z_{ij} = \delta_{ij} + \epsilon(\delta_{i, j+N_{\hop}} + \delta_{i,j-N_{\hop}}),
\end{equation}
where $j+N_{\hop}$ and $j-N_{\hop}$ are taken modulo $N$.  Using the fact that the perturbation involves cyclic permutation matrices $\sigma(\pm N_{\hop})$, we can do a power series expansion to extract $Z^{-1/2}$:
\begin{align}
Z^{-1/2} & = (I + \epsilon(\sigma(N_{\hop}) + \sigma(-N_{\hop}))^{-1/2} \nonumber \\
& = I - \frac{1}{2} \epsilon( \sigma(N_{\hop}) + \sigma(-N_{\hop})) +\frac{3}{8} \epsilon^2(2I + \sigma(2 N_{\hop}) + \sigma(-2N_{\hop})),
\end{align}
where $I$ is the identity matrix.  To order $\epsilon^2$, we thus arrive at \Eq{CyclicFieldRedef},
\begin{equation}
Z^{-1/2}_{ij} = \delta_{ij} - \frac12 \epsilon  \left ( \delta_{i, j+N_{\hop}} + \delta_{i,j-N_{\hop}} \right ) + \frac38 \epsilon^2 \left ( 2\delta_{ij} + \delta_{i, j+2N_{\hop}} + \delta_{i,j-2N_{\hop}} \right ).
\tag{\ref{CyclicFieldRedef}}
\end{equation}

The key property of the matrix $Z^{-1/2}$ is that its diagonal elements are order-$\epsilon^2$, so that to order $\epsilon$, the local terms (\ref{4ptLocal}) are odd in $\pi_i$, and hence do not contribute to singlet scattering. In fact, this property is much more general, and holds for any moose model, cyclic or linear, with or without a symmetry that sets all decay constants equal. To see this, note that a general local kinetic matrix is a diagonal matrix $D$, where the entries correspond to the different values of the decay constants $f_i$ for each link. After the addition of nonlocal perturbations, the kinetic matrix is of the form
\be
Z = D + \epsilon M,
\ee
where $M$ is a symmetric matrix with zero entries along the diagonal, since for $N_{\hop} < N$ the nonlocal terms only contain derivatives of the form $\partial \pi_i \partial \pi_j$ for $i \neq j$. Appealing to perturbation theory once again, the unperturbed eigenstates of $Z$ are simply the $\pi_i$, and the perturbed eigenvalues receive \emph{no} first-order contribution since $M$ has no diagonal entries. The first-order eigenstates are $\pi_i + \epsilon \sum_{j \neq i} c_{ij} \pi_j$, where $c_{ij}$ are some numerical coefficients which are not important for the discussion. The field redefinition matrix to first order is then
\be
Z^{-1/2} = P D^{-1/2} P^{-1}
\ee
where $D$ is the unperturbed kinetic matrix and $P$ is the matrix of eigenstates, with $P_{ij}= \delta_{ij} + \epsilon c_{ij}$. Then $P^{-1}_{ij} = \delta_{ij} - \epsilon c_{ij}$ to first order, and direct calculation shows that the diagonal entries of $Z^{-1/2}$ are only affected at order $\epsilon^2$.

\subsection{First-order eigenvector coefficients}
\label{app:FirstOrderEigenvector}

\begin{figure}[t]
\begin{center}
\includegraphics[scale=0.45]{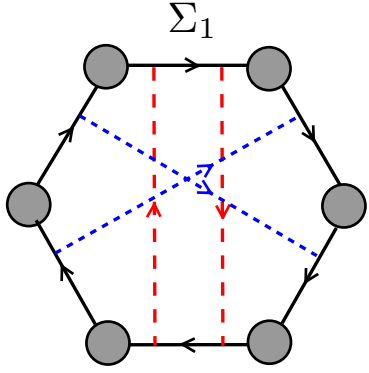}
\caption{Nonlocal terms contributing to $\langle S_1 | \T_1 | N_{\hop} \rangle$ for $N = 6$, $N_{\hop} = 3$.  Such terms must contain at least two factors of $\pi_1$; red (long-dashed) lines correspond to terms containing $\pi_1 \partial \pi_1$ and blue (short-dashed) lines contain $(\pi_1)^2$.  For large $N_{\rm hop}$, the blue lines give the dominant contribution to $\langle S_1 | \T_1 | N_{\hop} \rangle$.}
\label{6Site3Hop}
\end{center}
\end{figure}

To derive the first-order eigenvector shift \Eq{eq:FirstOrderEigenvec}, we need to compute $\langle 0 | \T_1 | \pm N_{\hop} \rangle$, which gets contributions from the Feynman diagrams in \Fig{fig:FirstOrder}.  Here we outline the calculation of these contributions. The first diagram comes from local terms on each site $i$, which contributes to the matrix elements $\langle \pi_i^a \pi_i^a | \T_1 | \pi_i^b \pi^b_{i \pm N_{\hop}} \rangle$ after order-$\epsilon$ field redefinitions from \Eq{CyclicFieldRedef}; this piece is subleading since it does not depend on $N_{\hop}$.

Turning to the second Feynman diagram involving the nonlocal term directly, there are two kinds of contributions to the 4-point terms containing two factors of $\pi_i$, as shown in \Fig{6Site3Hop} for the case $N = 6$, $N_{\hop} = 3$, and $i = 1$.  The red (long-dashed) lines represent terms which start or end on link $i$ and contain $\pi_i \partial \pi_i$:
\be
\label{EndsOn1}
 D \Sigma_i \cdots \Sigma_{i + N_{\hop} } D \Sigma_{i + N_{\hop}}^\dagger \cdots \Sigma_i^\dagger, \qquad D \Sigma_{i - N_{\hop}} \cdots \Sigma_i D\Sigma_i^\dagger \cdots  \Sigma_{i - N_{\hop}}, \\
\ee
The blue (short-dashed) lines represents terms which skip over link $i$ as they traverse the diagram and contain $(\pi_i)^2$:
\be
\label{HopsOver1}
D \Sigma_p \cdots \Sigma_i \cdots \Sigma_{p+N_{\hop}} D \Sigma_{p+N_{\hop}}^\dagger \cdots \Sigma_i^\dagger \cdots \Sigma_p^\dagger \hspace{2mm} (p = i - N_\hop + 1, \dots, i-1).
\ee
For any $N_{\hop} < N$, there are exactly two red lines, but there are $N_{\hop} - 1$ blue lines of the form (\ref{HopsOver1}); these latter terms dominate and give a scaling of $N_{\rm hop}$.  

Combining both Feynman diagrams, the matrix elements are
\be
\label{T1Nhop}
\langle 0 | \T_1 | \pm N_{\hop} \rangle = \frac{\epsilon s}{2f^2} \left( N_{\hop} - \frac12 \right).
\ee
From \Eq{CyclicFirstOrderEigenvec}, the first-order shifts in the eigenvectors are
\begin{equation}
c_{N_{\hop}}^{(1)} = c_{-N_{\hop}}^{(1)} = \epsilon (2N_{\hop} -1).
\end{equation}
This formula still holds for the exceptional case $N_{\hop} = N/2$, provided we write it as
\begin{equation}
c_{q}^{(1)} \equiv \frac{\langle 0  | \T_1 | q \rangle}{\lambda^{(0)}} = \epsilon (2N_{\hop} -1) (\delta_{q, N_{\hop}} + \delta_{q, -N_{\hop}})
\tag{\ref{eq:FirstOrderEigenvec}}
\end{equation}
as in the text, where the Kronecker deltas take into account the fact that $N_{\hop} \equiv -N_{\hop}$ contributes an extra factor of 2 to the amplitude.

\subsection{Second-order scattering matrix}
\label{app:SecondOrder}

The remaining ingredient to calculate the second-order eigenvalue shift $\lambda^{(2)}$ from \Sec{sec:NLDiagnostic} is $\langle 0 | \T_2 | 0 \rangle$.  As shown in \Figs{fig:SecondOrderFourPoint}{fig:SecondOrderThreePoint}, there are three contributions to the order-$\epsilon^2$ matrix elements:
\begin{itemize}
\item $\langle 0   | \T_2^{(1)} | 0 \rangle$: 4-point local terms after order-$\epsilon^2$ field redefinitions;\footnote{Note that these include contributions both from a single $\epsilon^2$ redefinition, and two $\epsilon$ redefinitions on two external legs; see the first two diagrams of \Fig{fig:SecondOrderFourPoint}.}
\item $\langle 0  | \T_2^{(2)} | 0 \rangle$: 4-point $\mathcal{O}_{\NL}$ terms after order-$\epsilon$ field redefinitions;
\item $\langle 0  | \T_2^{(3)} | 0 \rangle$: 3-point diagrams with no field redefinitions.
\end{itemize}
The calculation of these matrix elements is straightforward, although cumbersome:
\begin{align}
\label{T2Local}
\langle 0  | \T_2^{(1)} | 0 \rangle & = \frac{\epsilon^2 s}{f^2} \times \left\{ \begin{array}{lr}
1, & N_{\hop} \neq N/2,\\
2 ,        & N_{\hop} = N/2.
\end{array}
\right. \\
\label{T24pt}
 \langle 0 | \T_2^{(2)} |0 \rangle & =  -\frac{2\epsilon^2 s}{f^2} N_{\hop} \times \left\{ \begin{array}{lr}
  1, & N_{\hop} \neq N/2,  \\
 2 ,        & N_{\hop} = N/2.
\end{array}
\right. \\
\label{T23pt}
\langle 0 | \T_2^{(3)} |0 \rangle & = \frac{\epsilon^2 s}{2f^2} \times \left\{ \begin{array}{lr}
3N_{\hop} - 1, & N_{\hop} < N/2,  \\
3N_{\hop} -3 ,        & N_{\hop} \geq N/2.
\end{array}
\right.
\end{align}
The only noteworthy feature is the different behavior for $N_{\hop} \geq N/2$. The case \linebreak $N_{\hop} = N/2$ is special for the first two pieces because terms which contain derivatives separated by $N_{\hop}$ links contribute to 4-point diagrams, and these terms are duplicated either in field redefinitions (where shifting a link index forward or backward by $N/2$ lands on the same link) or in $\mathcal{O}_{\NL}$ itself.  Similarly, in the third piece, the case $N_{\hop} > N/2$ is special because the terms in the sum over $k$ in \Eq{3ptNL} contribute twice to the relevant matrix elements.

Summing over these three contributions, we find the subleading piece
 \begin{equation}
 \langle 0 | \T_2 | 0 \rangle = \frac{\epsilon^2 s}{2f^2} \times \left\{ \begin{array}{lr}
1 - N_{\hop}, & N_{\hop} < N/2,   \\
1 -5N_{\hop}, & N_{\hop} = N/2, \\
-1 - N_{\hop},        & N_{\hop} > N/2,
\end{array}
\right.
\end{equation}
which is always zero or negative. Including the $|\langle 0 | \T_1| q| \rangle|^2$ terms gives \Eq{SecondOrderShift},\footnote{The extra factor of 2 in the leading coefficient for $N_{\hop} = N/2$ is due to the fact that the sum over $q$ contains only \emph{one} matrix element, which is doubled compared to the case $N_{\hop} \neq N/2$; see the comment below \Eq{eq:FirstOrderEigenvec}. Squaring this gives a factor of $2^2 = 4$, compared to the single factor of 2 from summing $|\langle 0 | \T_1 | N_{\hop} \rangle|^2$ + $|\langle 0 | \T_1 | -N_{\hop} \rangle|^2$.}
\begin{align}
\lambda^{(2)} & = \langle 0 | \T_2 | 0 \rangle +   \sum_{q=1}^{N-1} \frac{ |\langle 0 | \T_1 | q \rangle|^2}{\lambda^{(0)}} \nonumber \\
& =  \epsilon^2 \, \lambda^{(0)} \times \left\{ \begin{array}{lr}
8N_{\hop}^2 - 10N_{\hop} + 4, & \quad N_{\hop} < N/2, \\
16N_{\hop}^2 - 26N_{\hop} + 6, & \quad N_{\hop} = N/2,  \\
8N_{\hop}^2 - 10N_{\hop} ,        & \quad N_{\hop} > N/2. 
\end{array}
\right.
\tag{\ref{SecondOrderShift}}
\end{align}
Plugging \Eq{SecondOrderShift} into \Eq{SWaveUnitarity} gives
\begin{equation}
\Lambda_{\NL} - \Lambda_{\rm local}  = -\epsilon^2 \Lambda_{\rm local} \times \left\{ \begin{array}{lr}
4N_{\hop}^2 - 5N_{\hop} + 2, & \quad N_{\hop} < N/2, \\
 8N_{\hop}^2 - 13N_{\hop} + 3, & \quad N_{\hop} = N/2,  \\
 4N_{\hop}^2 - 5N_{\hop} ,        & \quad N_{\hop} > N/2. 
\end{array}
\right.
\end{equation}
Note that the $\Lambda_{\NL} < \Lambda_{\rm local}$ for any $0 < N_{\hop} < N$, so the subleading corrections do not change the fact that $\R$ is maximized at $\epsilon = 0$.

\subsection{SU($\boldmath{n}$) moose}
\label{app:SUn}

It is straightforward to generalize the arguments in \Sec{sec:LocalScatteringMatrix} from an $\SU(2)$ gauge symmetry to an $\SUn$ gauge symmetry.  The global $\SUn$ symmetry of the moose means that the eigenvalues and eigenvectors of the scattering matrix must fall into representations of \linebreak Adj $\otimes$ Adj, which contains a unique copy of the trivial representation from the map \linebreak $\mathfrak{su}(n) \times \mathfrak{su}(n) \to \mathbb{C}$ given by the Killing form.\footnote{For $n > 3$, in addition to the singlet there are six other irreducible representations contained in \linebreak Adj $\otimes$ Adj, four of which have nonzero scattering amplitudes. See \Ref{Chivukula:1992gi} for details.} The corresponding eigenchannel for the largest eigenvalue of link $j$ is again the gauge singlet  \cite{Cahn:1991xf},
\begin{equation}
\label{SingletVec}
|S_j \rangle \equiv \frac{1}{\sqrt{2 (n^2-1)}} \sum_{a=1}^{n^2-1} |\pi_j^a \pi_j^a \rangle.
\end{equation}
The largest eigenvalue is
\be
\lambda_{\rm max} = \frac{n}{8} \frac{s}{f^2},
\ee
which implies a scale of unitarity violation\footnote{This agrees with the results in \Ref{Chang:2003vs} after a redefinition of the decay constant $f \to f/\sqrt{8}$.}
\be
\Lambda = 8\sqrt{\frac{\pi}{n}} f.
\ee
The definition of $\mbar$ does not depend on $n$, so for the local moose we have simply
\be
\label{LocalSUnRatio}
\R_{\rm{local} \, \SUn} = 4\sqrt{2} \sqrt{\frac{\pi}{n}}.
\ee
 
Turning now to the nonlocal moose, the definition of the $|q\rangle$ singlet states \eq{CyclicInvtStates} only requires a trivial modification for general $\SUn$,
\begin{equation}
\label{CyclicInvtStatesSUn}
| q \rangle_{\SUn}  \equiv \frac{1}{\sqrt{N}} \sum_{i = 1}^{N} \sum_{a = 1}^{n^2-1}\frac{1}{\sqrt{2(n^2-1)}} |  \pi_i^a \, \pi_{i + q}^a \rangle, \qquad q = 0, 1, \dots, N-1.
\end{equation}
The only parts of \Sec{sec:NSiteNL} which depended on the gauge group being SU(2) were the calculations of the exact numerical coefficients of the first-order eigenvector $c_i^{(1)}$, and the subleading corrections in $N_{\hop}$ of $\lambda^{(2)}$. For general $\SUn$, the following are still true:
\begin{itemize}
\item The first-order eigenvalue shift still vanishes;
\item The eigenvector shift $c^{(1)}$ is still proportional to $N_{\hop}$ because of the sum over intermediate states in the last line of \Eq{4ptNL};
\item The zeroth-order eigenvalue $\lambda^{(0)} = \frac{n}{8} \, \frac{s}{f^2}$ is still positive.
\end{itemize}
Thus, the eigenvalue shift will still be proportional to $\epsilon^2 N_{\hop}^2$ with a positive coefficient, up to subleading corrections in $N_{\hop}$. Depending on the particular form of the group theory factors for $\SUn$, the subleading corrections $\langle 0 | \T_2 | 0 \rangle$ may dominate and change the sign of $\lambda^{(2)}$ for small $N_{\hop}$, but for large enough ratios of $\ell/R$, maximizing $\R$ will still imply locality.

\bibliographystyle{JHEP}
\bibliography{MooseLocalBib}{}

\end{document}